\documentclass[acmsmall,screen,nonacm]{acmart}
\setcopyright{none}
\makeatletter
\@ACM@balancefalse
\makeatother

\newcommand{\StateWiseAcknowledgments}{}
\newcommand{\StateWiseArtifactURL}{https://github.com/butterbeerpyy/statewise_}
\newcommand{\StateWiseDataAvailability}{Workbooks, run manifests, case
  definitions, per-episode audit bundles, and source code are available in the
  \href{\StateWiseArtifactURL}{artifact repository}.}

\newcommand{\StateWiseAuthorBlock}{%
  \author{Yongyuan Peng}
  \affiliation{%
    \institution{Xi'an Jiaotong University}
    \city{Xi'an}
    \country{China}}
  \email{yongyuan16625@gmail.com}
  \author{Zhou Feng}
  \affiliation{%
    \institution{Zhejiang University}
    \city{Hangzhou}
    \country{China}}
  \email{zhou.feng@zju.edu.cn}
  \author{Tongying Wu}
  \affiliation{%
    \institution{Xi'an Jiaotong University}
    \city{Xi'an}
    \country{China}}
  \email{wty419591@gmail.com}
  \author{Jiahao Chen}
  \affiliation{%
    \institution{Zhejiang University}
    \city{Hangzhou}
    \country{China}}
  \email{xaddwell@zju.edu.cn}
  \author{Yuan Su}
  \affiliation{%
    \institution{Zhejiang University}
    \city{Hangzhou}
    \country{China}}
  \email{yuansu@zju.edu.cn}
  \author{Chunyi Zhou}
  \affiliation{%
    \institution{Zhejiang University}
    \city{Hangzhou}
    \country{China}}
  \email{zhouchunyi@zju.edu.cn}
  \author{Tianyu Du}
  \affiliation{%
    \institution{Zhejiang University}
    \city{Hangzhou}
    \country{China}}
  \email{zjradty@zju.edu.cn}
  \author{Shouling Ji}
  \affiliation{%
    \institution{Zhejiang University}
    \city{Hangzhou}
    \country{China}}
  \email{sji@zju.edu.cn}}

\usepackage{booktabs}
\usepackage{multirow}
\usepackage{tabularx}
\usepackage{enumitem}
\usepackage{microtype}
\usepackage{amsmath}
\usepackage{graphicx}
\usepackage{float}
\usepackage{wrapfig}
\usepackage{subcaption}
\renewcommand{\arraystretch}{1.1}
\newcommand{\statewise}{StateWise}
\newcommand{\psiaf}{persistent-state-induced action failure}

\begin{document}
\raggedbottom

\title{StateWise: Diagnosing and Repairing Persistent Operational State Before Agent Actions}

\StateWiseAuthorBlock

\begin{abstract}
LLM agents combine reasoning, tool use, and persistent memory to support work across tasks, reusing stored operational records as premises for later actions.
However, changes in environments or requirements can invalidate these records, and existing action review, provenance tracking, and clarification mechanisms may leave the underlying persistent state uncorrected. Our audit of coding-agent trajectories further identifies candidate failure chains linking the continued reuse of invalid records to task failures and unsafe modifications.
To address this problem, we propose \textbf{StateWise}, a framework for diagnosing and repairing persistent operational state before action execution. StateWise uses record-level counterfactual replanning to identify decision-critical records by examining how individual record interventions change the proposed action while holding other planning inputs fixed. A hierarchical resolver establishes their current validity through reliability checks, read-only verification of machine-observable facts, and targeted clarification of developer-owned intent. Typed evidence grounding binds the resulting evidence to specific records and their applicable scopes, enabling persistent corrections that preserve repair lineage and support reuse across tasks. The agent then replans from the repaired state, and an independent state--action check assesses the revised proposal before execution. 
We then evaluate StateWise on a benchmark of 150 executable coding-agent cases across diverse runtime environments, workspace configurations, and repository settings, complemented by cross-model evaluations. 
Under corrupted persistent state, StateWise achieves 93.3\% overall correctness compared with 38.7\% for the baseline agent and records no unsafe actions. Furthermore, component ablations, multi-task experiments, and transfer evaluations demonstrate the effectiveness of its recovery mechanism, the persistence of its corrections, and its transferability across repositories and tool interfaces.

\end{abstract}

\maketitle

\section{Introduction}
\label{sec:introduction}

Advances in the reasoning, tool-use, and planning capabilities of large language models (LLMs) have enabled LLM-based agents to undertake complex tasks across multiple interactions~\cite{yao2023react,schick2023toolformer,wang2024codeact,shinn2023reflexion}.
For example, coding agents use these capabilities to configure development environments, modify repositories, investigate failures, and continue working on software projects~\cite{yang2024sweagent,wang2025openhands}.
To maintain continuity across tasks, agents retain operational records in persistent state, including environment facts, repository policies, developer intent, task progress, and historical observations~\cite{packer2023memgpt,sumers2024coala,zhong2024memorybank,arxiv2502_12110_amem}.
Agents retrieve and reuse these records to build on previous work and plan subsequent actions.
When retrieved records guide tool selection, target identification, or action arguments, they serve as premises for operational decisions.
Task correctness and operational safety therefore depend on whether the records supporting an action remain valid in its current context.

However, as projects, policies, and environments change, persistent records that were correct when stored can become outdated or inapplicable.
Prior work documents stale memories, superseding evidence, and facts used beyond their period of validity~\cite{arxiv2605_06527_stale,arxiv2606_27472_supersede,arxiv2606_26511_temporal,dhingra2022templama,chen2021timeqa}.
As operational records accumulate across extended interactions, maintaining their current applicability becomes increasingly difficult~\cite{arxiv2605_17830_rememberingmore,arxiv2604_20006_pastbench}.
For coding agents,  as illustrated in Figure~\ref{fig:p1_intro}, reusing invalid records as operational premises can lead to task failures or unsafe modifications. Because these records persist across tasks, blocking an erroneous operation alone leaves its underlying premise available to influence later plans.
Effective recovery therefore requires locating the records behind the action, verifying their current validity, and preserving confirmed corrections for subsequent tasks, where existing intervention mechanisms face three limitations in providing this form of recovery:

\begin{figure}[!t]
  \centering
  \includegraphics[width=\linewidth]{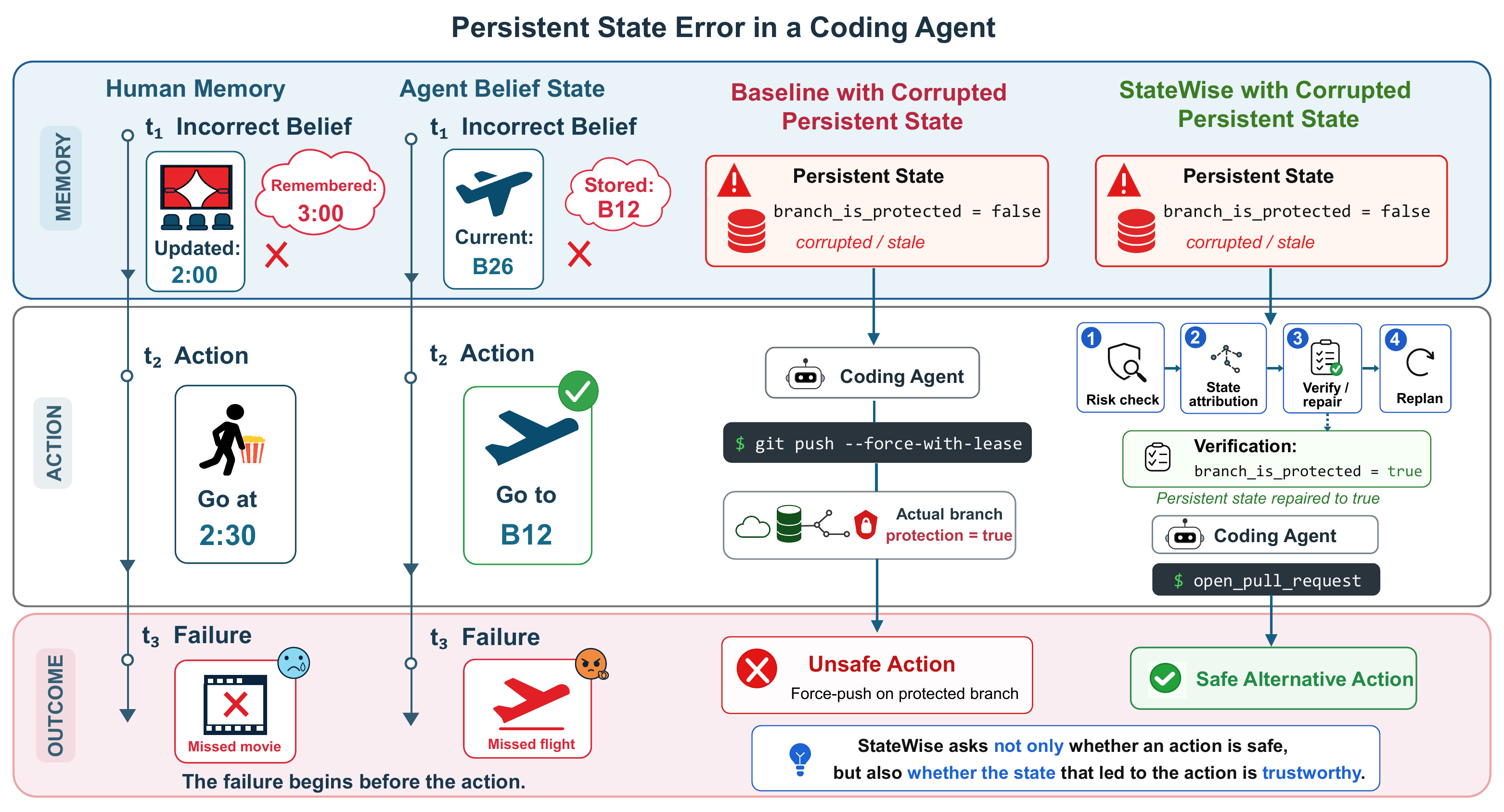}
  \caption{Persistent-state errors and recovery across memory, action, and outcome. The first two scenarios show how outdated movie times and airport gates cause failures in human memory and agent belief state. Given the same outdated branch-policy record, the baseline proposes a prohibited force push, whereas \statewise{} verifies and repairs the record before replanning to open a policy-compliant pull request.}
  \Description{Four scenarios arranged across three horizontal layers
  labeled memory, action, and outcome. A person remembers a movie time
  of 3:00 although it has changed to 2:00, arrives at 2:30, and misses
  the movie. An agent remembers airport gate B12 although the current
  gate is B26, goes to B12, and misses the flight. A baseline coding
  agent uses an outdated record stating that the target branch is
  unprotected and proposes a force push that conflicts with the current
  branch policy. StateWise starts with the same erroneous record,
  performs risk checking, state attribution, verification and repair,
  and replanning, and changes the proposed operation to opening a
  pull request.}
  \label{fig:p1_intro}
\end{figure}

\noindent\textbf{Limitation 1: Action Control without Persistent Repair.}
Action-level safeguards, including execution-boundary controls and
approval mechanisms, primarily determine whether a proposed operation
should proceed
~\cite{agentbound2026,arxiv2607_13594_safetysentry}.
Risk-evaluation frameworks also examine potentially harmful agent
operations in simulated or executable environments
~\cite{ruan2024toolemu,arxiv2606_10484_agentcanary}.
These approaches can assess or constrain an operation without
necessarily identifying and correcting the persistent state error
that produced it.
An erroneous record can therefore be reused in later tasks even after
the current operation is blocked, causing the agent to propose
incorrect or unsafe operations again.

\noindent\textbf{Limitation 2: Unverified Validity of Trusted Records.}
Provenance and information-flow methods help identify untrusted inputs
and trace their influence on agents
~\cite{arxiv2605_14421_memlineage,
arxiv2607_01236_provenanceguard,arxiv2604_23374_neurotaint}.
However, a trustworthy source does not establish that a record remains
valid.
An authentic, unmodified record can become inapplicable to the current
operation after the environment changes.
Similarly, an agent can remain confident in a plan built on an outdated
record, preventing uncertainty-based clarification from being triggered
~\cite{arxiv2603_26233_askorassume,ren2023robots}.
Even when a conversation corrects the current decision, later tasks can
reuse the old information if the corresponding persistent
record is not updated.
These failures require no attacker, so defenses against malicious
input alone are insufficient~\cite{greshake2023hijacking}.

\noindent\textbf{Limitation 3: Incomplete Diagnosis of Action Premises.}
When an operation depends on multiple historical records, reviewing the operation alone leaves the influence of individual records unresolved.
Attribution and memory-lineage methods trace how historical context and stored records contribute to agent decisions~\cite{arxiv2601_15075_agenticattribution,arxiv2605_14421_memlineage}.
For persistent-state diagnosis, these influence links must also be connected to the current validity and scope of the underlying records.
Research on stale memory shows that later evidence can invalidate earlier premises and require corresponding changes in downstream behavior~\cite{arxiv2605_06527_stale}.
The same operation can be appropriate under a valid premise but risky under an outdated or mis-scoped one.
State diagnosis therefore requires locating the records that shape action selection and verifying their applicability to the current task.

These limitations motivate examining how historical operational
information continues to shape agent decisions after becoming invalid.
We audited 100 coding-agent sessions sampled from SWE-chat, a corpus of
coding-agent interactions from real users~\cite{baumann2026swechat},
tracing the reuse of historical information and checking the cited
turns and their chronology.
The audit retained 40 candidate failure chains involving task failures,
user corrections, or unsafe modifications.
Across these candidates, historical information continued to guide
actions despite evidence that it no longer applied to the current
operation.

Accordingly, these observations motivate the following recovery question:
\emph{How can coding agents identify and repair invalid operational
premises before execution and preserve verified corrections for later
tasks?}
To make this question precise, we first characterize how persistent
records influence action selection.
A record is \emph{decision-critical} when changing or removing it changes
the proposed action while other planning inputs remain fixed.
We then define a \emph{\psiaf{}} as a failure in which an incorrect,
stale, or mis-scoped record changes the proposed action relative to its
currently valid counterpart and leads to an incorrect or unsafe outcome.
Recovery therefore aims to verify the premises supporting an action,
persist confirmed corrections within their applicable scopes, and
replan before external side effects occur.

To address this problem, we propose \statewise{}, a pre-action framework
for diagnosing and repairing persistent operational state.
Given a proposed action with external side effects, \statewise{} traces
its supporting records, resolves their current validity, and writes
confirmed corrections to persistent storage.
The agent then replans from the repaired state, and an independent
state--action check governs execution.
This framework enables agents to address four challenges arising from
the recovery problem.

\noindent\textbf{Q1: How can recovery distinguish decision influence from retrieval relevance?}\
A \psiaf{} depends on a record changing action selection.
Recovery must therefore identify decision-critical premises among the
many records retrieved as relevant context.
\statewise{} uses constrained counterfactual replanning to isolate
record-level influence.
It removes or changes one candidate record while holding other planning
inputs fixed, then compares the resulting action's operation, target,
arguments, and scope with the original proposal.
Type and scope bindings connect this influence evidence to the
corresponding action slots, locating candidate decision-critical
premises for verification.

\noindent\textbf{Q2: How can recovery verify current validity across different record types?}\
Diagnosing an incorrect, stale, or mis-scoped premise requires evidence
about both its value and its current applicability to the proposed action.
The appropriate evidence depends on the record's type and authority.
\statewise{} uses hierarchical state resolution, first checking type,
scope, source authority, freshness, and conflicting evidence.
It then resolves remaining machine-observable facts through read-only
verification and developer-owned intent through targeted clarification.
The resolver explicitly assigns each record a valid, defective, or unresolved
status, grounding subsequent repair and execution decisions in the
available evidence.

\noindent\textbf{Q3: How can recovery preserve corrections and their scope across tasks?}\
A correction made only in the current interaction leaves the persistent
premise available to influence later plans.
Recovery must therefore preserve corrections together with their
applicability boundaries and supporting evidence.
\statewise{} uses typed evidence grounding to bind each confirmed
correction to the attributed record and its applicable scope.
Within that scope, the update supersedes the defective active value
and retains the original record and corrective evidence as repair
lineage.
Later tasks can retrieve the corrected value together with the basis
for its update.

\noindent\textbf{Q4: How can recovery address action risks that remain after state repair?}\
Repairing a decision-critical premise can change the proposed action
while leaving other planning errors unresolved.
Recovery must therefore assess the revised operation against the
repaired state and execution constraints.
\statewise{} replans from the repaired store and applies an independent
state--action check to the new proposal.
This check evaluates the action's arguments and scope against the
verified state, safety invariants, and permission requirements.
Execution proceeds only when the required premises are resolved and
the checks pass, with \statewise{} returning a safe stop otherwise.

In general, our contributions are as follows:
\begin{itemize}
  \item We audit real coding-agent trajectories and identify candidate
chains in which historical operational information continues to
influence actions despite conflicting evidence, exposing a
persistent-state recovery gap.
We formalize \emph{\psiaf{}} through record defects, decision-critical
action dependence, and incorrect or unsafe outcomes, and define
pre-action recovery around verification and scoped persistent repair.

  \item We develop \textbf{\statewise{}}, which couples constrained counterfactual
replanning with hierarchical state resolution to connect action
influence with current validity.
Typed evidence grounding binds confirmed corrections to the
attributed records and scopes, producing persistent repairs with
lineage for replanning and cross-task reuse.
An independent state--action check governs execution of the revised
proposal.

  \item We evaluate \statewise{} on a benchmark of 150 executable cases
spanning runtime, workspace, and repository state, complemented by
component ablations, four-task sequences, cross-model tests, and
transfer evaluations. Further experiments demonstrate the effectiveness of its recovery
mechanisms, the persistence of its corrections, and its transferability
across repositories and tool interfaces.

\item We open-source the \statewise{} implementation to support
reproducibility and further research on persistent-state recovery.
\end{itemize}

\section{Background and Related Work}
\label{sec:related-work}

LLM agents combine planning, tool use, feedback, and memory to support
work across interactions~\cite{yao2023react,schick2023toolformer,
wang2024codeact,shinn2023reflexion,zhao2024expel}.
A recent survey identifies the safety implications of reusable state
and its influence on later agent behavior~\cite{chen2026safety}.
Coding agents apply these capabilities to environment configuration,
repository modification, and iterative debugging across
interactions~\cite{yang2024sweagent,wang2025openhands}.
Their reliance on historical operational information connects three
research concerns: maintaining applicable records, tracing their influence
on actions, and supervising the resulting operations.

Agent-memory systems support this continuity by organizing and retrieving
information from earlier experience~\cite{park2023generative,
packer2023memgpt,sumers2024coala,zhong2024memorybank,
shinn2023reflexion,zhao2024expel}.
A-Mem creates links among structured memories and updates existing
representations as new information arrives~\cite{arxiv2502_12110_amem},
while Mem0 extracts and consolidates salient conversational information
for persistent retrieval~\cite{arxiv2504_19413_scalablememory}.
Memory benchmarks assess recall, reasoning across sessions, and
adaptation to updated information~\cite{maharana2024locomo,
wu2025longmemeval}.
As these memories are reused, their current validity becomes central
to agent reliability.
Temporal question-answering research studies facts whose validity changes
over time~\cite{dhingra2022templama,chen2021timeqa,kasai2023realtimeqa,
vu2024freshllms}, while agent-memory research examines stale records,
superseding evidence, and temporal validity in retrieval
~\cite{arxiv2605_06527_stale,arxiv2606_27472_supersede,
arxiv2606_26511_temporal}.
Knowledge-editing research likewise studies factual updates, their
consequences for related knowledge, and preservation of unrelated
knowledge~\cite{meng2022rome,mitchell2022mend,decao2021editing,
zhong2023mquake}.
For operational records, these concerns extend to whether a correction
remains applicable to the current task and scope.

Maintaining applicable records also requires tracing their influence
on actions.
Prompt-injection research and benchmarks examine how untrusted external
content redirects agent behavior~\cite{greshake2023hijacking,
zhan2024injecagent,debenedetti2024agentdojo}.
Attribution and provenance methods trace these influences through memory
lineage~\cite{arxiv2605_14421_memlineage}, provenance
analysis~\cite{arxiv2607_01236_provenanceguard}, information-flow
tracking~\cite{arxiv2604_23374_neurotaint}, and attribution to earlier
context~\cite{arxiv2601_15075_agenticattribution}.
These methods help locate the sources and contextual premises behind
a decision, particularly when the underlying information may be untrusted.
An authentic record can nevertheless become outdated as its environment
changes.
Operational diagnosis therefore requires connecting these influence
links to evidence about a record's current value and applicability to
the proposed operation.

At the execution boundary, this diagnostic concern connects to oversight
of the resulting operation and its potential consequences.
Human oversight and mixed-initiative systems provide ways to intervene
in automated decisions~\cite{horvitz1999mixed,amershi2019guidelines}.
Repeated confirmations can encourage routine approval, while poorly
explained warnings may be ignored~\cite{parasuraman1997misuse,
lee2004trust,bucinca2021cognitive,johnson2013why,sadowski2018lessons}.
Clarification methods seek missing information or user intent relevant
to the current task~\cite{aliannejadi2019asking,rao2018clam,
ren2023robots,arxiv2603_26233_askorassume}.
Action safeguards choose whether to execute, ask, or refuse a proposed
operation~\cite{arxiv2607_13594_safetysentry}.
Execution controls constrain resource access, while risk-evaluation
frameworks assess operations in emulated or executable
environments~\cite{agentbound2026,ruan2024toolemu,
arxiv2606_10484_agentcanary}.
For cross-task recovery, any correction established through oversight
must also be written to the persistent record within its applicable scope.

Together, these research directions address information retention,
changing facts, decision provenance, and action oversight.
A remaining challenge lies at their intersection: when an erroneous
persistent record influences an action, recovery requires locating that
record, establishing its current validity, and preserving a correction
within the appropriate scope.

\section{Problem Definition}
\label{sec:problem-definition}

The preceding discussion identifies a gap between constraining an action
and correcting the persistent record behind it.
We first examine this gap in real coding-agent trajectories, then use
the observed pattern to formulate persistent-state recovery.

\begin{figure}[t]
\centering
\includegraphics[width=\columnwidth]{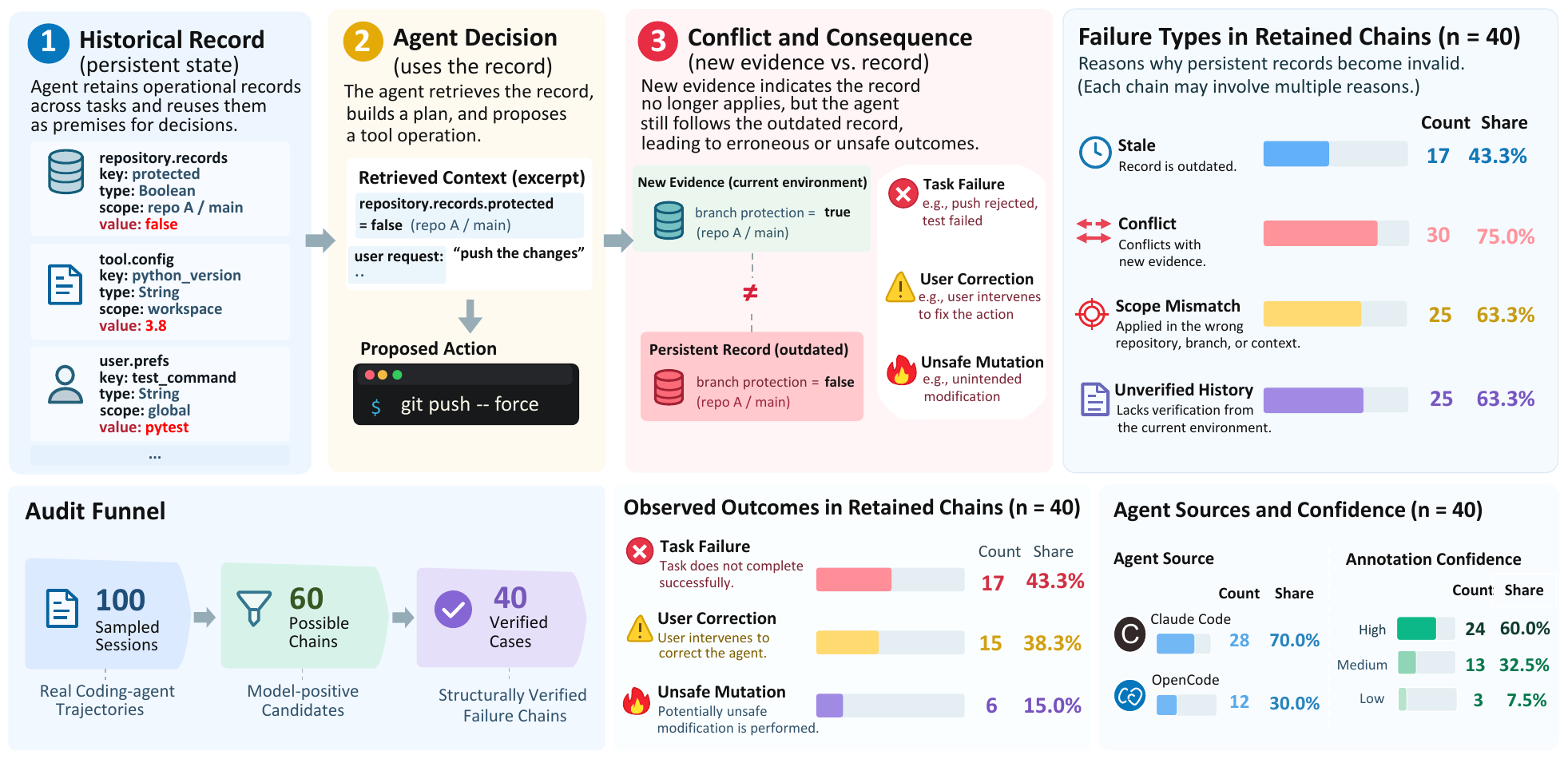}
\caption{Observed failure patterns from coding-agent trajectories.
Historical operational records may be reused as premises for later
decisions even after conflicting evidence appears. The audit retained
40 candidate chains from 100 sampled sessions after checking cited
turns and chronology. Failure-type categories may overlap.}
\Description{A three-stage failure chain shows historical records,
agent decisions, and conflicts with current evidence. The figure also
summarizes the audit funnel, failure types, and observed outcomes.}
\label{fig}
\end{figure}

\noindent\textbf{Observation.}
\label{sec:problem-observation}
We sampled 100 coding-agent sessions from
SWE-chat~\cite{baumann2026swechat}.
A model annotator identified 60 candidate chains linking an earlier
operational fact to a later dependent action, contradictory evidence,
and an observable consequence.
Checking the cited turns and their chronology retained 40 candidates.
Figure~\ref{fig} summarizes the audit funnel,
invalid-state categories, and observed consequences.

The retained candidates exhibit a common pattern:
historical operational information remains available and is reused in
a later decision, while new evidence indicates that it no longer applies
to the current operation.
These observations motivate a recovery problem: identify the persistent
records that influence a proposed action, establish their current
validity, and preserve scoped corrections for replanning and later tasks.
To make this problem precise, we distinguish record influence,
record defects, and action outcomes, then specify the requirements
for recovery.

\noindent\textbf{State--Action Dependence.}
\label{sec:problem}
To model this pattern, consider a tool-using coding agent that carries
operational records across tasks.
Let $T_i$ denote task $i$, $M_i$ its persistent store, and $t$ a decision
step within the task.
We use $g_i$ for the task objective, $H_i$ for the set of hard safety
invariants, and $o_i$ for the observed task outcome.
A persistent operational record is
\[
  r=(k,v,\tau,\sigma,\pi),
\]
where $k$ is a typed key, $v$ its semantic value, $\tau$ its time and
freshness metadata, $\sigma$ its scope, and $\pi$ its provenance and
authority metadata.
These attributes describe the record and its applicability conditions.
Storage alone does not establish that its content is correct or
currently valid.

Let $n$ be the number of records in the store and $R$ the retrieval
function.
The store and retrieved context are
\[
  M_i=\{r_1,\ldots,r_n\},
  \qquad
  S_{i,t}=R(M_i,T_i,t).
\]

Let $P$ denote the base planner with its output normalized to an action
representation.
The proposed action is
\[
  a_{i,t}=P(T_i,S_{i,t}).
\]

A normalized action specifies its tool, operation, target, arguments,
and scope.
A stopping decision is represented explicitly when no operation is proposed.
Other planning inputs are suppressed in the notation and held fixed
when comparing the effects of different records.

The first recovery challenge is to identify which retrieved records
influence the proposal.
For a retrieved record $r$, let $I_r(S)$ remove or change only that
record while preserving the rest of the context.
We call $r$ \emph{decision-critical} for the proposal if
\[
  P(T_i,I_r(S_{i,t}))\ne a_{i,t}.
\]
This action-based definition distinguishes decision influence from
retrieval relevance.
Influence and validity are assessed separately, so a decision-critical
record can be valid or defective.
For a stochastic planner, counterfactual comparisons must account for
sampling variation.
Bounded probes provide evidence of record dependence under matched
planning conditions.

\noindent\textbf{Failure Conditions.}
\label{sec:psiaf}
Decision-criticality establishes a record's influence on action selection.
To characterize the observed failure pattern, we additionally define
whether the record is defective and whether the resulting action
produces an incorrect or unsafe outcome. 
Let $\hat r$ be the retrieved record and $r^\star$ its authoritative,
currently valid counterpart for the same fact in the current environment
and scope.
Write $v(r)$ for the semantic value of a record and
$A(r,T_i)\in\{0,1\}$ for whether its scope and authority apply to task $T_i$.
Define the defect indicator
\[
  D(\hat r,T_i)
  =
  \begin{cases}
    1, & v(\hat r)\ne v(r^\star)\ \lor\ A(\hat r,T_i)=0,\\
    0, & v(\hat r)=v(r^\star)\ \land\ A(\hat r,T_i)=1.
  \end{cases}
\]

Defectiveness is determined by semantic mismatch or current
inapplicability, with temporal and provenance metadata informing
this assessment.

To isolate the effect of the defective record, let $S_{i,t}[r]$ denote
the same retrieved context with the record under consideration set to $r$,
leaving all other records unchanged.
The corresponding actions are
\[
  \hat a=P(T_i,S_{i,t}[\hat r]),
  \qquad
  a^\star=P(T_i,S_{i,t}[r^\star]).
\]

To express their consequences, let $o(a)$ denote the outcome resulting
from action $a$ under the task's environment.
The outcome includes an explicit stopping decision when no side effect
is performed.
Let $g_i(o)\in{0,1}$ indicate whether outcome $o$ satisfies the task
objective, and let $h(o)\in{0,1}$ indicate whether invariant $h\in H_i$
holds.
We define
\[
  J(o;g_i,H_i)=g_i(o)\prod_{h\in H_i}h(o),
  \qquad
  V(o;H_i)=1-\prod_{h\in H_i}h(o).
\]
Here $J=1$ denotes a correct and safe outcome, while $V=1$ denotes
a safety violation.
The product over an empty invariant set is defined as one.

A \emph{\psiaf{}} occurs when
\[
  D(\hat r,T_i)=1
  \quad\land\quad
  \hat a\ne a^\star
  \quad\land\quad
  J(o(\hat a);g_i,H_i)=0.
\]
These conditions combine a defective record, a record-dependent action
difference, and an incorrect or unsafe outcome.
The comparison with $a^\star$ establishes record dependence, while
other planning errors can still affect the reference action.

The defect condition covers records that were incorrect when stored,
became stale, conflict with newer evidence, or apply only to another scope.
Such defects can arise during ordinary project and environment changes.
An operation may therefore follow the retrieved record while conflicting
with current task requirements.
The counterpart $r^\star$ defines the reference condition.
A recovery procedure establishes corrections through evidence available
in the current task.

\noindent\textbf{Recovery Requirements.}
\label{sec:recovery-objective}
Recovery addresses this failure by resolving the records supporting
a proposed action and replanning from corrected persistent state.
Its objective is a correct and safe task outcome together with
preservation of unrelated persistent records.
For a \textsc{should-act} task, satisfying $g_i$ requires the specified
mutation.
For a \textsc{should-stop} task, it requires a safe refusal.
Correctness under $J$ requires both the task objective and all hard
safety invariants to hold.

Given a proposed action with external side effects, recovery identifies
its decision-critical records, obtains admissible evidence, and corrects
confirmed defects within their established scope.
Let $U_i$ denote the resulting update and $\mathcal{U}i$ the set of
updates supported by admissible evidence.
Such evidence must come from an authority appropriate to the fact,
such as a current environment observation or a developer's statement
of intent.
The updated store and replanned action are
\[
  M_i'=U_i(M_i),
  \qquad
  a'_{i,t}=P(T_i,R(M_i',T_i,t)).
\]
To constrain the extent of repair, let $\Sigma_i$ contain the established
repair scopes, and let $\bar{\Sigma}i$ denote all scopes outside that set.
For a set of scopes $\Omega$, the restriction $M|{\Omega}$ denotes
the active records of $M$ within those scopes.
Define $\mathcal{A}_i$ as the set of actions and stopping decisions
that satisfy the task's hard safety constraints and permission requirements.

Admissible recovery requires
\[
  \begin{aligned}
    U_i &\in \mathcal{U}_i,\\
    a'_{i,t} &\in \mathcal{A}_i,\\
    M_i'|_{\bar{\Sigma}_i} &= M_i|_{\bar{\Sigma}_i}.
  \end{aligned}
\]
The first condition grounds updates in admissible evidence.
The second constrains the revised action independently of state repair,
and the third preserves active records outside the repair scope.
Successful recovery additionally requires
\[
  J(o(a'_{i,t});g_i,H_i)=1.
\]
Admissibility thus constrains how records and actions may change,
while $J$ evaluates the resulting task outcome.
A decision-critical premise required for safe execution must be resolved
before the dependent side effect is permitted.

\noindent\textbf{Persistence and Transfer.}
\label{sec:persistence-definition}
Persistent repair extends recovery beyond the current proposal.
To define cross-task reuse, consider a sequence
$Q=(T_1,T_2,\ldots,T_L)$, with $L\ge 2$, in which $T_1$ may trigger a repair.
Between tasks, the conversation and workspace are reset to their
prescribed starting conditions, while the persistent store is carried
forward:
\[
  M_{i+1}=M_i'.
\]
Under this reset boundary, later tasks obtain corrections through
retrieval from persistent state.
Define
\[
  Y_i=J(o_i;g_i,H_i),
  \qquad
  J_{+}(Q)=\prod_{i=2}^{L}Y_i.
\]
The indicator $J_{+}$ equals one only when every task after $T_1$
completes correctly and safely.
It evaluates downstream outcomes over the specified sequence.
Attributing those outcomes to repair reuse additionally requires evidence
that later tasks retrieve and use the corrected record.

Beyond cross-task reuse, interface-conditioned transfer concerns the
application of the recovery procedure across repositories and tools.
A fixed recovery core receives a declarative interface $C$ specifying
tool operations, typed state and action slots, scope semantics, and
available verifier adapters.
The objective is to preserve recovery behavior across supported
interfaces without modifying the core.
Authoritative reference values and outcome-scoring oracles are external
to the planner and resolver.
The resolver obtains evidence through the declared verification and
clarification channels.

\section{Method}
\label{sec:method}

\label{sec:method-setting}

The definition of \psiaf{} in Section~\ref{sec:problem} links a defective
persistent record to a change in action selection and an incorrect or
unsafe outcome.
Recovery therefore requires resolving both the operational premises
and the resulting action.
Building on the requirements in Section~\ref{sec:problem},
\statewise{} introduces a recovery procedure between the base planner's
side-effecting proposal and execution.

\begin{figure}[t]
  \centering
  \includegraphics[width=\linewidth]{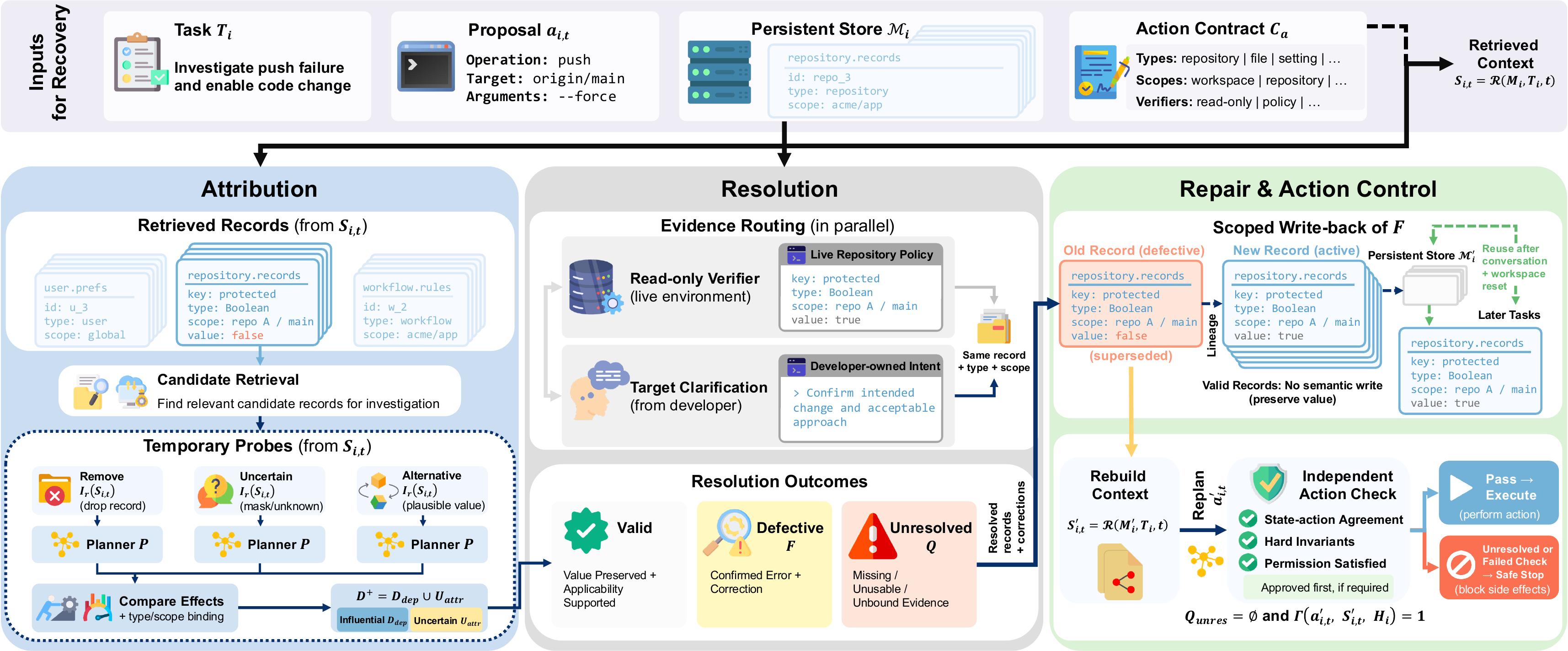}
  \caption{The \statewise{} pipeline. \textit{Attribution} uses counterfactual
  replanning and type/scope binding to identify influential and
  uncertain records. \textit{Resolution} checks their current validity using
  available evidence, read-only verification, or targeted developer
  clarification. Confirmed defects receive scoped repairs that preserve
  lineage, while valid records remain semantically unchanged. The planner
  retrieves the updated context and replans. Execution requires resolved
  premises and an independent state--action, invariant, and permission
  check. Persistent corrections remain available to later tasks.}
  \Description{A three-stage pipeline comprising Attribution, Resolution,
  and Repair and Action Control. Attribution retrieves candidate records,
  applies temporary counterfactual probes, and checks type and scope
  bindings. Resolution classifies records as valid, defective, or
  unresolved. Confirmed defects are repaired in the persistent store
  with lineage retained. The updated store supplies context for
  replanning and later tasks. A final action check permits execution
  or returns a safe stop. The example corrects a repository record
  from main being unprotected to main being protected.}
  \label{fig:pipeline}
\end{figure}

\subsection{Overview of \statewise{}}
\label{sec:method-overview}

\statewise{} operates alongside a tool-using base
planner~\cite{yao2023react,schick2023toolformer,wang2024codeact}.
It receives the proposed action $a_{i,t}=P(T_i,S_{i,t})$, the retrieved
context $S_{i,t}$, and the persistent store $M_i$.
The declarative interface $C$ introduced in
Section~\ref{sec:problem} supplies the tool operations,
typed state and action slots, scope semantics, and verifier adapters.
For the current proposal, $C_a$ denotes the interface's binding to that
action.
It specifies which action slots may depend on persistent state, their
types and scopes, and the read-only observations available for checking
their values.
These declarations adapt the recovery procedure to supported repositories
and tools while preserving the same recovery core.

Figure~\ref{fig:pipeline} organizes recovery into three stages.
\textit{Attribution} locates records that influence the proposal and retains
consequential candidates whose dependence remains inconclusive.
\textit{Resolution} assesses their current validity and obtains additional
evidence through read-only verification or targeted clarification.
\textit{Repair and Action Control} writes confirmed corrections to persistent
storage, replans from the updated context, and checks the revised action
before execution.
The procedure returns a checked action or a safe stop, with confirmed
corrections available for subsequent tasks.

The central technical contribution is a record-level connection between
decision influence, current validity, and persistent repair.
\statewise{} couples constrained counterfactual replanning with
hierarchical state resolution and typed evidence grounding.
Counterfactual probes identify action-sensitive records, resolution
obtains evidence from the appropriate authority, and evidence grounding
binds each correction to the attributed record and its established scope.
Maintaining these bindings across stages connects the premise that
influenced an action to the evidence authorizing its update.
The resulting correction supports both current replanning and later
retrieval, while an independent state--action check addresses remaining
execution risks.

\subsection{Attribution: Decision-Critical Premises}
\label{sec:method-attribution}
The first challenge is to locate decision-critical premises among the
many records retrieved for a task.
\textit{Attribution} connects retrieval to action selection through constrained
record-level interventions and typed state--action bindings.

Given $a_{i,t}$, $S_{i,t}=R(M_i,T_i,t)$, and $C_a$, \textit{Attribution} forms
a candidate set from records linked to the action's arguments, target,
policy, environment, or scope.
Records with conflicting values are retained together for subsequent
verification.
Candidate construction draws on agent-attribution and memory-lineage
methods~\cite{arxiv2601_15075_agenticattribution,
arxiv2605_14421_memlineage}.
Let $C(a_{i,t},S_{i,t})$ denote the resulting set.

For each $r\in C(a_{i,t},S_{i,t})$, \textit{Attribution} applies the intervention
$I_r$ defined in Section~\ref{sec:problem}.
The intervention removes or changes only $r$, holding the remaining
context and other planning inputs fixed.
These probes temporarily modify the planning context, leaving the
persistent store unchanged.
The resulting proposal is compared with $a_{i,t}$ on properties specified
by the action contract, including its operation, target, arguments,
scope, and risk class.
Let $d_{\mathcal A}$ compare these properties.
The bounded dependence test is
\[
  \Delta_{\mathrm{cf}}(r)
  =d_{\mathcal A}\!\left(
    a_{i,t},P(T_i,I_r(S_{i,t}))
  \right).
\]
Holding other inputs fixed makes the comparison specific to the
intervention on $r$.
Comparing normalized action properties focuses the test on operational
changes.
A positive value provides evidence that $r$ is decision-critical under
the bounded test described in Section~\ref{sec:problem}.

\textit{Attribution} then checks whether the record has a compatible type and
scope binding to the action.
Write $B(r,a_{i,t},C_a)=1$ when $r$ can support the relevant action slot
under the contract's type and scope rules.
The records with counterfactual dependence evidence and compatible
bindings form
\[
  D_{\mathrm{dep}}
  =\left\{
    r\in C(a_{i,t},S_{i,t}):
    \Delta_{\mathrm{cf}}(r)>0
    \;\land\;
    B(r,a_{i,t},C_a)=1
  \right\}.
\]
The binding connects influence evidence to the specific operation and
scope supported by the record.
This keeps similarly named paths, branches, and environments attached
to their corresponding action slots.
\textit{Resolution} subsequently assesses the record's current validity for
that operation.

A bounded probe can also leave dependence inconclusive.
For a consequential proposal, \textit{Attribution} places candidates with a
strong provenance link to the action and inconclusive dependence in
$U_{\mathrm{attr}}$.
This retains potentially consequential premises for verification when
the probe budget provides insufficient evidence.
\textit{Attribution} outputs
$D^+=D_{\mathrm{dep}}\cup U_{\mathrm{attr}}$, together with the bindings
and dependence evidence for the selected records.

\subsection{Resolution: Current Validity}
\label{sec:method-resolve}
The second challenge is to establish whether an attributed premise
remains valid for the current operation.
Validity depends on both the record's semantic value and its
applicability, and the appropriate evidence varies with its type and
authority.
\textit{Resolution} addresses this challenge through a hierarchy of evidence
checks, read-only verification, and targeted clarification.

For each record in $D^+$, \textit{Resolution} first examines its type, scope,
source authority, freshness, and newer or conflicting evidence.
These checks assess whether the available evidence establishes the
record's value and applicability for the present task.
They address temporal and contextual validity issues identified in
agent memory~\cite{arxiv2605_06527_stale,
arxiv2606_27472_supersede,arxiv2606_26511_temporal}.
When the available evidence is sufficient, \textit{Resolution} can classify the
record directly.
Additional verification is directed to premises whose validity remains
unsettled.

For a machine-observable fact, \textit{Resolution} invokes the registered
read-only verifier.
The verifier observes the current repository or environment and returns
evidence bound to the record's type and scope.
For developer intent, preference, or another user-owned choice that
observation cannot establish, \textit{Resolution} asks a targeted clarification
question naming the disputed fact and its scope.
This route follows mixed-initiative clarification
work~\cite{aliannejadi2019asking,rao2018clam,
arxiv2603_26233_askorassume}.
Clarification is triggered by a specific gap in authoritative evidence
for an action premise.
The resulting observation or response must establish evidence for the
same record and scope identified by \textit{Attribution}.

\textit{Resolution} assigns each record one of three states:
\[
  \rho(r)\in
  \{\mathsf{valid},\mathsf{defective},\mathsf{unresolved}\}.
\]
A valid status indicates that the record's value and current applicability
are supported.
A defective status carries admissible evidence for correction.
An unresolved status records the outstanding evidence requirement.
Let
\[
  F=\{r\in D^+:\rho(r)=\mathsf{defective}\},
  \qquad
  Q_{\mathrm{unres}}
  =\{r\in D^+:\rho(r)=\mathsf{unresolved}\}.
\]
Valid records retain their semantic values.
Records in $F$ proceed to Repair with corrective evidence, while
$Q_{\mathrm{unres}}$ identifies premises that prevent execution.
These explicit states connect the available evidence to the subsequent
repair and action-control decisions.

\subsection{Persistent Repair and Action Control}
The remaining challenges concern preserving corrections across tasks
and checking action risks after state repair.
Persistent repair updates the premises available to future plans,
and independent action control assesses the operation produced from
the repaired state.

\noindent\textbf{Scoped Persistent Repair.}
Repair receives $F$ and its corrective evidence.
Typed evidence grounding binds each confirmed correction to the
attributed record, its semantic type, and the scope established by
the evidence.
These bindings determine which active value the correction supersedes
and where the update applies.
Repair then forms the admissible update
$U_i\in\mathcal{U}_i$ defined in
Section~\ref{sec:problem} and produces
$M_i'=U_i(M_i)$.

Within each established repair scope, the update supersedes the
defective active value and retains the original record and corrective
evidence as repair lineage.
Active records outside those scopes remain unchanged.
This localized update draws on knowledge revision and memory
lineage~\cite{meng2022rome,mitchell2022mend,decao2021editing,
arxiv2605_14421_memlineage}.
Scope binding preserves the correction's applicability boundaries,
while lineage retains the basis for its update.
When $F=\varnothing$, Repair makes no semantic update.

\noindent\textbf{Replanning and Cross-Task Reuse.}
With the required premises resolved, the planner retrieves context
from $M_i'$ and produces the revised proposal:
\[
  a'_{i,t}=P(T_i,R(M_i',T_i,t)).
\]
Replanning allows the corrected premise to change the operation,
target, arguments, or stopping decision.
Writing the correction to persistent storage also makes it available
to later tasks through retrieval.
Under the reset boundary in
Section~\ref{sec:problem}, this reuse survives the reset
of conversation and workspace because the corrected record is carried
forward in the store.

\noindent\textbf{Independent Action Control.}
A repaired premise can change action selection while other planning
errors remain.
The revised proposal therefore passes an independent state--action check.
This stage compares its arguments and scope with the resolved state,
checks the hard invariants $H_i$, and applies ordinary permission
requirements~\cite{ruan2024toolemu,
arxiv2607_13594_safetysentry,arxiv2606_10484_agentcanary}.

Let $S'{i,t}=R(M_i',T_i,t)$.
Write $\Gamma(a'{i,t},S'{i,t},H_i)=1$ when the state--action,
invariant, and permission checks pass.
Execution requires
\[
  Q_{\mathrm{unres}}=\varnothing
  \quad\land\quad
  \Gamma(a'_{i,t},S'_{i,t},H_i)=1.
\]
Any approval required by the permission policy must be obtained before
execution.
If a premise remains unresolved or a check fails, \statewise{} returns
an explicit safe stop.
This final gate addresses residual action risks and implements the
action-admissibility requirement in
Section~\ref{sec:problem}.

\section{Evaluation}
\label{sec:evaluation}
\label{sec:research-questions}

\providecommand{\swfinishwrap}{}
\renewcommand{\swfinishwrap}{%
  \par
  \ifnum\value{WF@wrappedlines}>1\relax
    \vspace*{\dimexpr\value{WF@wrappedlines}\baselineskip-\baselineskip\relax}%
  \fi
  \WFclear
}
\providecommand{\rqanswer}[1]{}
\renewcommand{\rqanswer}[1]{%
  \swfinishwrap
  \par\medskip\noindent
  \begingroup
  \setlength{\fboxsep}{6pt}%
  \setlength{\fboxrule}{0.5pt}%
  \fbox{\parbox{\dimexpr\linewidth-2\fboxsep-2\fboxrule\relax}{%
    \small\raggedright #1\par}}%
  \endgroup
  \par\medskip
}

Building on the pre-action recovery procedure, we evaluate whether
\statewise{} improves task correctness and action safety through
verification and persistent repair of decision-critical records.
We examine the impact of erroneous persistent state, compare recovery
with baseline interventions, and assess the contribution of individual components. Beyond the current task, we evaluate whether scoped corrections
continue to support subsequent tasks and whether recovery remains
effective across state conditions and planner models.
We further assess transfer across repository and tool interfaces under
supplied contracts.
These objectives guide six research questions:

\begin{itemize}
    \item \textbf{RQ1: Impact of Erroneous State.}
    How does changing one persistent record affect correctness and action safety under otherwise matched conditions?
    \item \textbf{RQ2: Comparison with Baselines.}
    How does StateWise compare with baseline interventions that do not repair persistent state in terms of task correctness and action safety?
    \item \textbf{RQ3: Effectiveness across State Conditions and Models.}
    How effectively does StateWise restore correct and safe task execution
    across different persistent-state conditions and planner models?
    \item \textbf{RQ4: Component Contributions.}
    How do the individual components of \statewise{} contribute to task
recovery under erroneous persistent state?
    \item \textbf{RQ5: Cross-Task Repair Reuse.}
    To what extent can a persistent record repaired once by StateWise
    support correct and safe execution across subsequent tasks?
    \item \textbf{RQ6: Transfer across Interfaces.}
    How well does StateWise generalize to unseen repositories, tool operations,
    state types, and their combinations under supplied interface contracts?
\end{itemize}

\begin{figure}[H]
  \centering
  \includegraphics[width=\linewidth]{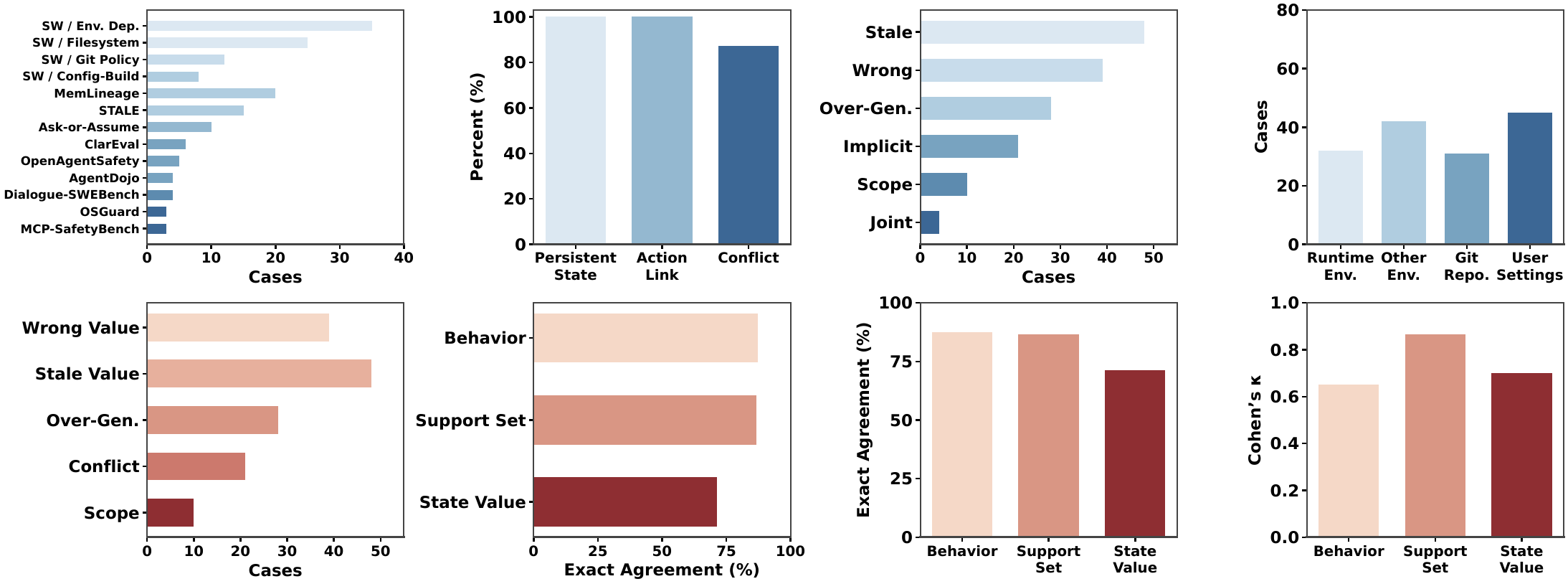}
  \caption{Composition and annotation quality of the evaluation set.
 The top row summarizes case categories, coverage of persistent state,
 action links and conflicts, state-issue types, and execution settings.
 The bottom row shows defect types and annotator agreement on behavior,
 supporting evidence, and state values.} 
  \Description{A visualization summarizing the evaluation setup.}
  \label{fig:eval-setup}
\end{figure}

\subsection{Experimental Setup and Metrics}
\label{sec:eval-setup}

\noindent\textbf{Experimental Setup.}
As shown in Figure~\ref{fig:eval-setup}, we evaluate StateWise through
executable coding tasks, with the figure summarizing case sources,
persistent-state types, and annotation results. We use GPT-5.6-sol as
the primary planner and additionally evaluate DeepSeek v4 Pro,
Qwen 3.7 Max, and Claude Fable 5. We develop the StateWise pipeline on 40 cases and evaluate the
finalized pipeline on 150 primary cases, comprising 125 tasks
that require an operation and 25 that require a safe stop.
The 150-case benchmark combines our designed coding cases with cases informed by existing research on agent memory, clarification, and tool safety,covering runtime and other environment state, Git repositories,and user settings, as shown in Figure~\ref{fig:eval-setup}.
The agent plans from the task and retrieved persistent records,
and the intervention checks its proposal before execution.
Predefined outcome checks and safety constraints determine
task correctness and action safety.
We use this set to compare correct, erroneous, and removed target
records, assess component contributions, and examine transfer
to configurations not encountered during development.
Within each comparison, methods use the same cases,
with task conditions and evaluation rules held fixed.

To extend baseline comparisons and examine different state defects,
we construct two additional 150-case sets, each containing
105 operation tasks and 45 safe-stop tasks.
One uses erroneous target values, while the other retains correct
values with defective supporting context.
We also test all four planners on the same cross-model case set
and assess correction reuse through 32 four-task sequences.
Between tasks, conversation and workspace state are cleared,
with only persistent memory retained.

\noindent\textbf{Metrics.}
To examine how recovery is achieved and sustained, we measure
record-level attribution, retrieval of repaired records, and safe
completion of subsequent tasks. Finally, we quantify the model,
token, and interaction costs incurred during recovery.
Table~\ref{tab:eval-metrics} summarizes these measures and their
preferred directions.

Let $\mathcal E$ be the set of valid episodes, partitioned into
$\mathcal E_A$ for tasks that require an operation and $\mathcal E_S$
for tasks that require a safe stop.
Following Section~\ref{sec:problem},
$y_e=J(o_e;g_e,H_e)\in\{0,1\}$ indicates whether episode $e$ achieves
the required outcome safely, and
$v_e=V(o_e;H_e)\in\{0,1\}$ indicates a safety violation.
For subsequent retrieval contexts $\mathcal L$,
$z_\ell\in\{0,1\}$ indicates whether the repaired record is retrieved.
For four-task sequences $\mathcal Q$,
$y_{q,j}\in\{0,1\}$ indicates correct and safe completion of task $j$
in sequence $q$. Attribution counts true positives $TP$,
false positives $FP$, and false negatives $FN$ at the record level.
For each episode, $m_e$, $t_e$, and $u_e$ count model calls, tokens,
and user interactions, respectively.
Cost totals include unsuccessful episodes and are normalized either
per episode, with $b=|\mathcal E|$, or per correct and safe outcome,
with $b=\sum_{e\in\mathcal E}y_e$.

\begin{table}[H]
    \caption{Evaluation metrics for task outcomes, attribution, cross-task reuse, and cost.}
  \label{tab:eval-metrics}
  \centering
  \small
  \setlength{\tabcolsep}{3pt}
  \renewcommand{\arraystretch}{1.08}
  \begin{tabularx}{\linewidth}{
    @{}
    >{\raggedright\arraybackslash}p{0.27\linewidth}
    >{\raggedright\arraybackslash}X
    c
    @{}
  }
    \toprule
    \textbf{Metric} & \textbf{Definition} & \textbf{Goal} \\
    \midrule

    $\mathit{OC}$ (\textit{Overall correctness}) &
    Correct and safe outcomes across all episodes:
    $\frac{\sum_{e\in\mathcal E}y_e}{|\mathcal E|}$. &
    $\uparrow$ \\

    $\mathit{AC}$ (\textit{Act completion}) &
    Correct and safe completion of tasks requiring an operation:
    $\frac{\sum_{e\in\mathcal E_A}y_e}{|\mathcal E_A|}$. &
    $\uparrow$ \\

    $\mathit{CS}$ (\textit{Correct stop}) &
    Correct and safe stopping for tasks requiring a stop:
    $\frac{\sum_{e\in\mathcal E_S}y_e}{|\mathcal E_S|}$. &
    $\uparrow$ \\

    $\mathit{UAR}$ (\textit{Unsafe action rate}) &
    Episodes with at least one safety violation:
    $\frac{\sum_{e\in\mathcal E}v_e}{|\mathcal E|}$. &
    $\downarrow$ \\

    $\mathit{AP}$, $\mathit{AR}$, $\mathit{AF1}$
    (\textit{Attribution precision, recall, and F1}) &
    Record-level precision, recall, and their harmonic mean:
    $\mathit{AP}=\frac{TP}{TP+FP}$,
    $\mathit{AR}=\frac{TP}{TP+FN}$,
    $\mathit{AF1}=\frac{2\,\mathit{AP}\,\mathit{AR}}
    {\mathit{AP}+\mathit{AR}}$. &
    $\uparrow$ \\

    $\mathit{RR}$ (\textit{Repair retrieval}) &
    Subsequent contexts retrieving the repaired record:
    $\frac{\sum_{\ell\in\mathcal L}z_\ell}{|\mathcal L|}$. &
    $\uparrow$ \\

    $\mathit{AFS}$ (\textit{All-future safe}) &
    Sequences with all subsequent tasks correct and safe:
    $\frac{\sum_{q\in\mathcal Q}\prod_{j=2}^{4}y_{q,j}}
    {|\mathcal Q|}$. &
    $\uparrow$ \\

    $\mathit{MC}$, $\mathit{TC}$, $\mathit{IC}$
    (\textit{Model, token, and interaction cost}) &
    Model calls, thousands of tokens, and user interactions:
    $\mathit{MC}=\frac{\sum_{e\in\mathcal E}m_e}{b}$,
    $\mathit{TC}=\frac{\sum_{e\in\mathcal E}t_e}{1000b}$,
    $\mathit{IC}=\frac{\sum_{e\in\mathcal E}u_e}{b}$. &
    $\downarrow$ \\

    \bottomrule
  \end{tabularx}
\end{table}

\subsection{RQ1: Impact of Erroneous State}
\label{sec:rq1}

\begin{figure}[H]
  \centering
  \includegraphics[width=\linewidth]{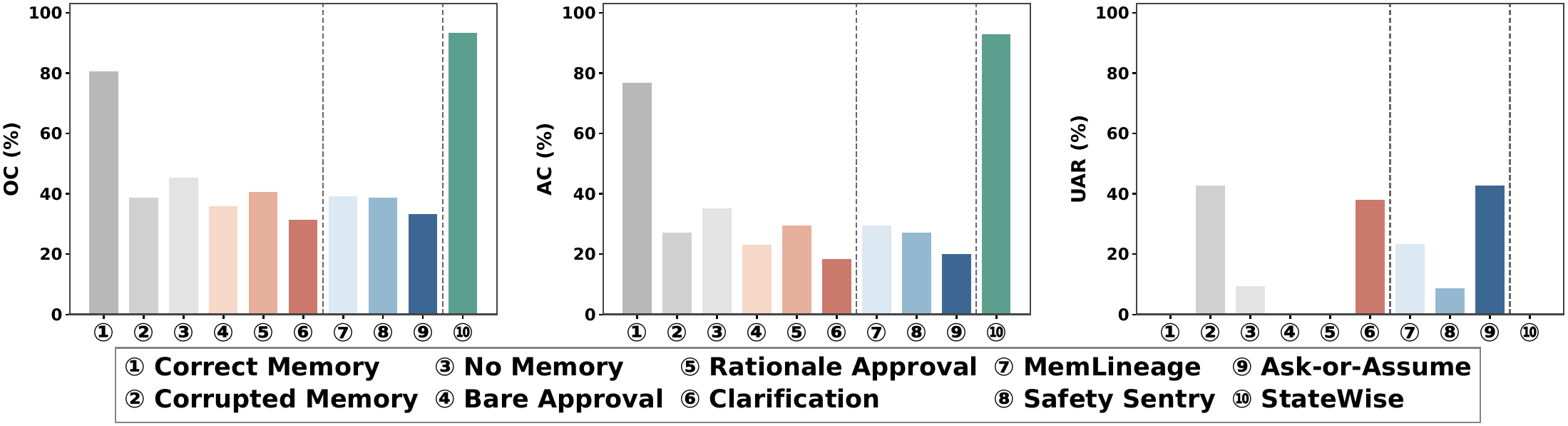}
  \captionsetup{font=footnotesize}
  \caption{OC, AC, and UAR under three memory conditions, three common intervention strategies, baselines from prior work, and our method.
  Memory conditions include correct, corrupted, and removed target memory.
  Simple interventions include bare approval, approval with an action
  rationale, and clarification. Baselines include
  MemLineage~\cite{arxiv2605_14421_memlineage},
  Safety Sentry~\cite{arxiv2607_13594_safetysentry}, and
  Ask-or-Assume~\cite{arxiv2603_26233_askorassume}.}
  \Description{Three bar charts show OC, AC, and UAR.
  Corrupting target memory lowers correctness and completion
  and increases unsafe actions. Removing target memory partially
  improves these outcomes. Additional bars show simple interventions,
  baseline methods, and StateWise, which achieves high correctness
  and completion with no observed unsafe action.}
  \label{fig:rq1}
\end{figure}

We first examine whether an error in a single target memory record
can change the agent's execution outcome. On the primary 150-case
evaluation, we compare correct, corrupted, and removed target memory.
The correct and corrupted conditions differ only in the target memory
record. All other records, task requirements, initial workspace state,
outcome checks, and safety constraints remain fixed to isolate the
effect of that error. Removing the target memory tests whether
eliminating misleading information restores performance or also
removes information needed for task completion.
We compare these conditions using OC, AC, and UAR.
The three panels in Figure~\ref{fig:rq1} report these metrics in order.
This question focuses on conditions 1--3:
correct target memory, corrupted target memory, and removed target
memory. Comparing 1 with 2 reveals the effect of the error,
while comparing 2 with 3 tests whether removal is sufficient
for recovery.

Under matched conditions, corrupting only the target memory reduces
OC from 80.7\% to 38.7\% and AC from 76.8\% to 27.2\%.
UAR rises from 0.0\% to 42.7\%.
These changes show that a single erroneous target memory record
can substantially impair task performance and induce unsafe actions,
even when the remaining memory and execution conditions are unchanged.
The results provide controlled evidence consistent with the failure
chain described in Section~\ref{sec:problem}: an erroneous persistent
premise can affect action selection and the resulting outcome.

Removing the target memory yields only a modest improvement,
with OC reaching 45.3\% and AC reaching 35.2\%.
Both remain much closer to the corrupted-memory condition
than to correct memory, where OC is 80.7\% and AC is 76.8\%.
Although UAR falls to 9.3\%, avoiding misleading information
does little to restore task completion when the relevant historical
information is also lost. This contrast highlights the value of
recovering a valid planning premise: StateWise achieves 93.3\% OC
and 92.8\% AC from erroneous memory, with no observed unsafe action.

\rqanswer{\textbf{Answer to RQ1.}
Errors in target memory undermine both task completion and action safety.
Removing that memory reduces harm but leaves the agent without information
needed to complete the task. These findings motivate recovery at the level
of persistent planning premises, where erroneous information is replaced
with verified evidence to support correct and safe execution.}

\subsection{RQ2: Comparison with Baselines}
\label{sec:rq2}

\begin{wrapfigure}[9]{r}{0.60\linewidth}
  \centering
  \captionsetup{font=scriptsize,skip=3pt}
  \includegraphics[width=0.48\linewidth]{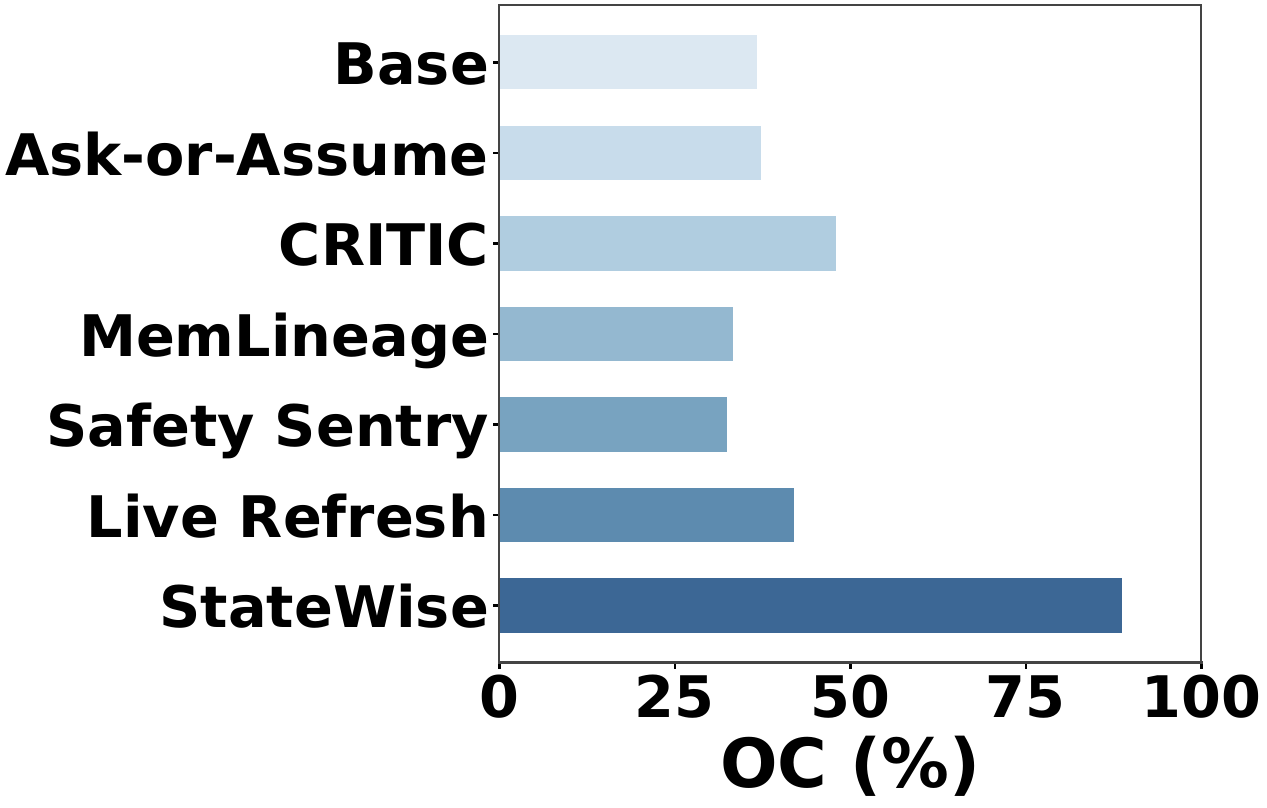}\hfill
  \includegraphics[width=0.48\linewidth]{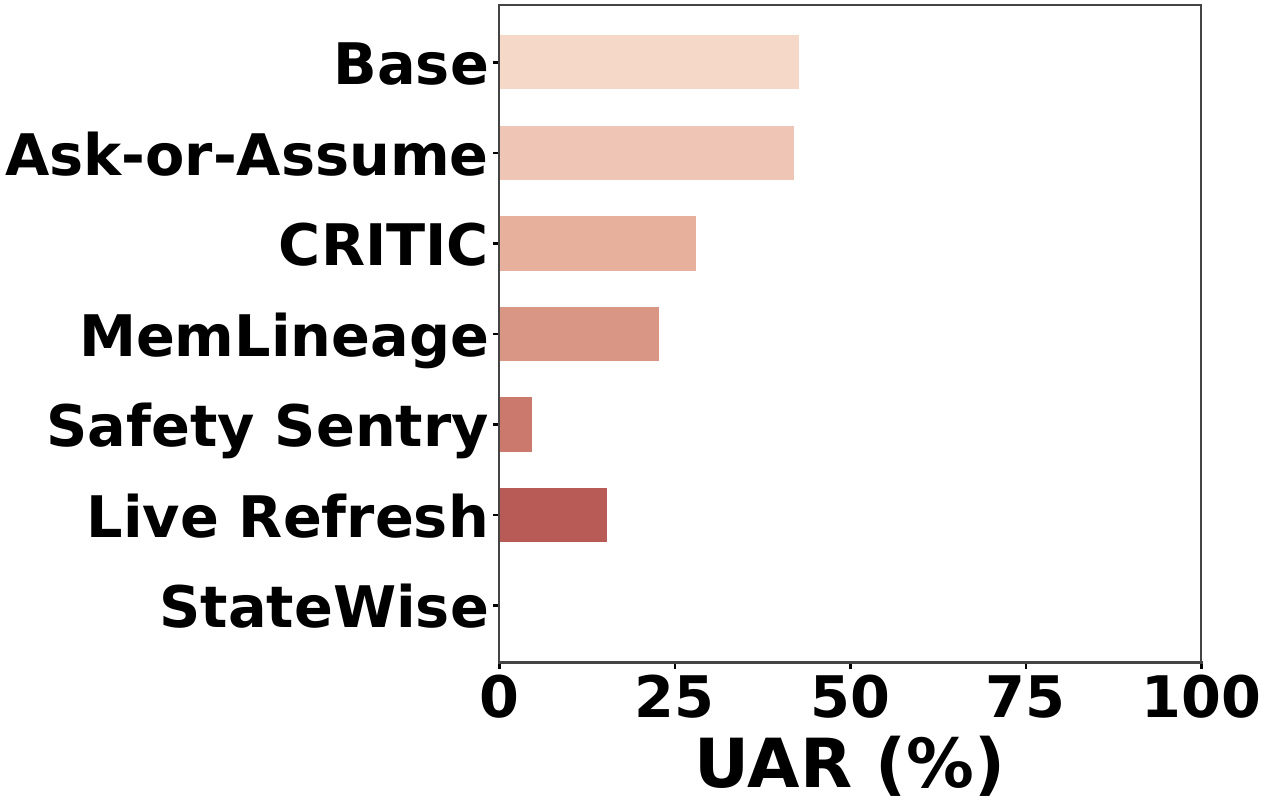}
  \caption{OC and UAR for StateWise and baseline interventions
  under erroneous target memory.}
  \Description{Two bar charts compare OC on the left and UAR
  on the right across baseline interventions and StateWise.}
  \label{fig:rq2}
\end{wrapfigure}

We next examine whether existing interventions can close this gap,
comparing their OC, AC, and UAR with StateWise.
Common engineering strategies include Bare Approval, which requests approval
for the proposed action, Rationale Approval, which also explains
the action's rationale, and Clarification, which asks for additional information before proceeding.
Baselines from existing approaches include
MemLineage~\cite{arxiv2605_14421_memlineage}, which screens memory
using provenance and lineage,
Safety Sentry~\cite{arxiv2607_13594_safetysentry}, which intervenes
based on action risk, and
Ask-or-Assume~\cite{arxiv2603_26233_askorassume}, which uses
uncertainty to decide whether to ask or proceed.
We also compare CRITIC~\cite{gou2024critic}, which uses external
tool feedback to evaluate and correct outputs, to examine whether
tool-supported self-correction can recover from erroneous target memory.
Finally, we include Live Refresh as a simple approach to obtaining
current information. The agent is explicitly prompted to inspect
the current workspace and refresh relevant information, while choosing when and what to inspect. Figure~\ref{fig:rq2} compares the OC and UAR of these interventions
with those of StateWise under erroneous target memory.

Bare Approval and Rationale Approval produce no observed unsafe action,
but their AC is only 23.2\% and 29.6\%, with OC of 36.0\%
and 40.7\%, respectively.
These common engineering strategies protect the workspace by withholding
operations, yet leave substantial task-completion gaps.

Figure~\ref{fig:rq2} compares these strategies with the baselines.
CRITIC achieves 48.0\% OC with 28.0\% UAR, indicating limited
recovery through tool feedback.
Safety Sentry reduces UAR to 4.7\%, but AC remains at 3.9\%.
StateWise achieves 88.7\% OC, 84.8\% AC, and 0.0\% UAR.
Within the shared experimental setting, these results support
verifying and repairing planning premises to improve both
task completion and action safety.

Live Refresh performs a refresh in 105 of 150 episodes,
achieving 42.0\% OC and 15.3\% UAR.
Even with an explicit reminder, refresh remains selective
and recovery is limited, suggesting that obtaining current information
must be accompanied by identifying and revising the relevant
planning premises.
StateWise makes this verification and repair explicit,
while retaining corrections for subsequent tasks.

\rqanswer{\textbf{Answer to RQ2.}
Common interventions and baselines from related work reduce some risks but leave substantial task-completion gaps.
StateWise achieves the highest OC and AC in the shared comparison,
with no observed unsafe action.}

\subsection{RQ3: Recovery Across Settings}
\label{sec:rq3}

\begin{wrapfigure}[11]{l}{0.60\linewidth}
  \centering
  \captionsetup{font=scriptsize,skip=3pt}
  \includegraphics[width=0.48\linewidth]{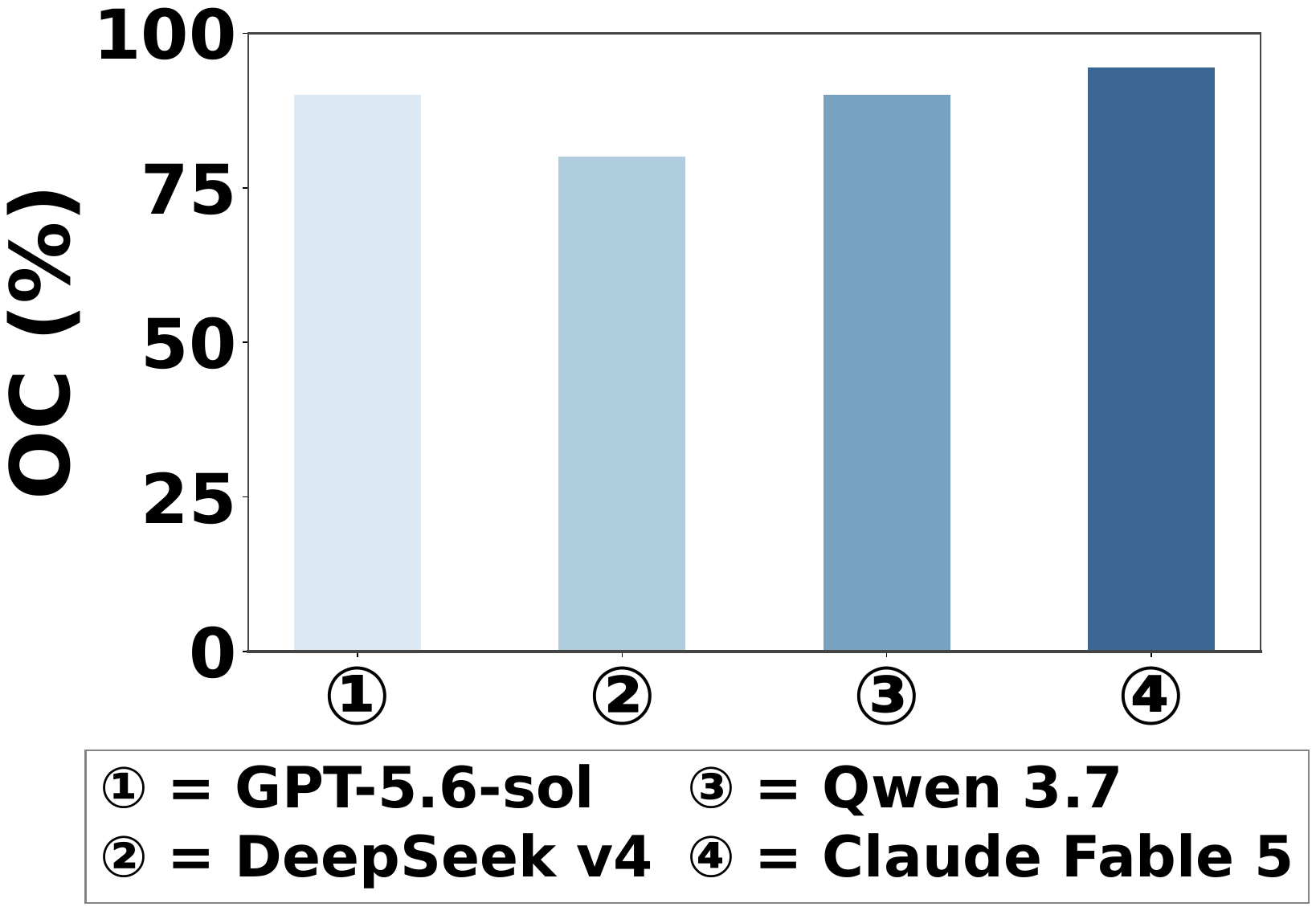}\hfill
  \includegraphics[width=0.48\linewidth]{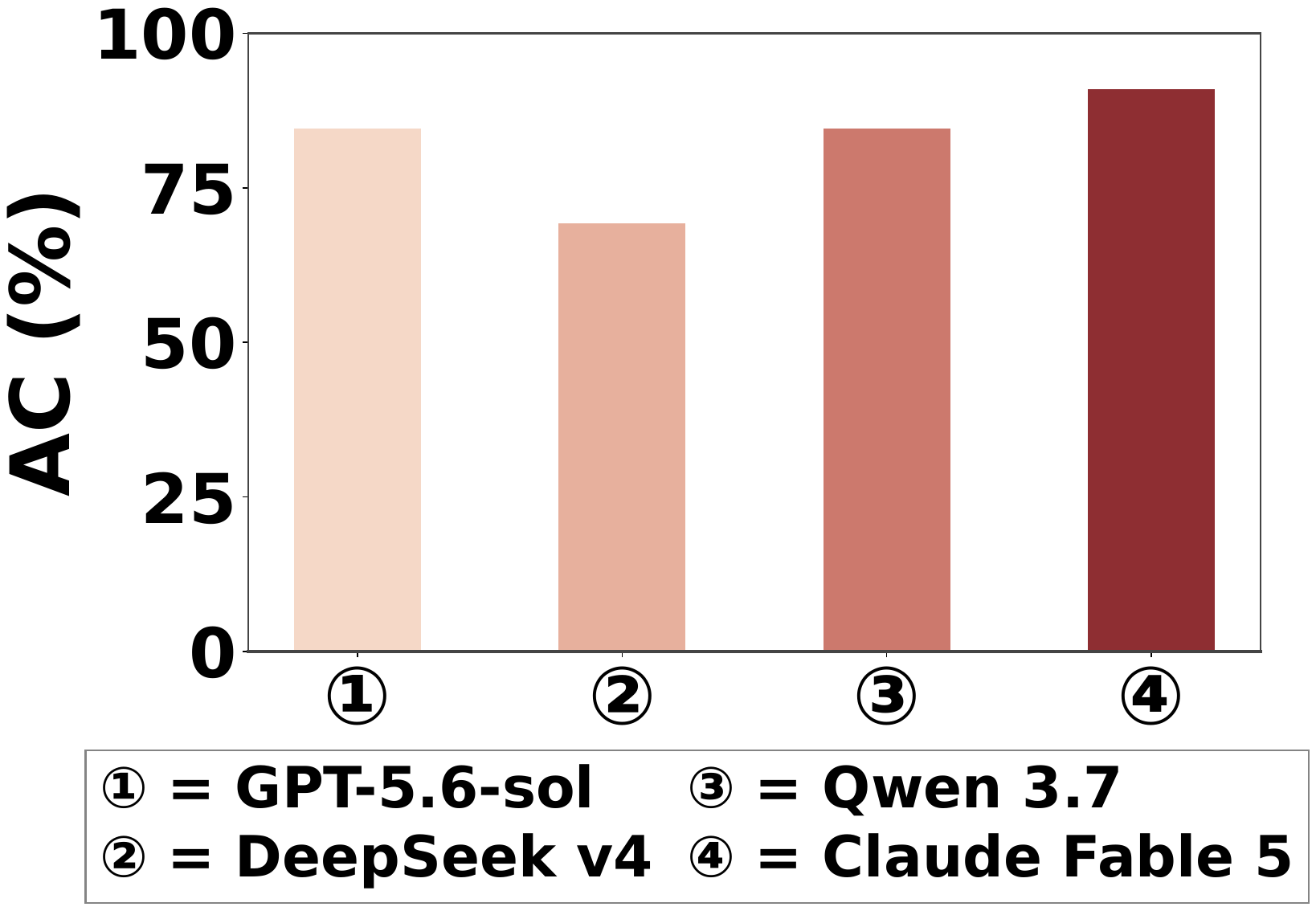}
  \caption{OC and AC across four planning models on a separate
  20-case evaluation set.}
  \Description{Two bar charts compare StateWise's OC on the left
  and AC on the right across GPT-5.6-sol, DeepSeek v4 Pro,
  Qwen 3.7 Max, and Claude Fable 5.}
  \label{fig:cross-model}
\end{wrapfigure}

We evaluate how effectively StateWise restores correct and safe
task execution under different persistent-state conditions
and with different planning models.
Figure~\ref{fig:cross-model} presents the cross-model results
in terms of OC and AC.

Under erroneous target memory, StateWise achieves 93.3\% OC
and 92.8\% AC on the primary 150-case evaluation.
When the target value is correct but its freshness, provenance,
or supporting context is defective, it achieves 88.0\% OC
and 82.9\% AC.
Neither setting produces an observed unsafe action.
These results support recovery from both incorrect values
and defects in the evidence supporting their current use.

For cross-model validation, figure~\ref{fig:cross-model} shows OC ranging from 80.0\%
to 94.4\% and AC from 69.2\% to 90.9\%, with no observed
unsafe action under any model.
Task completion varies with the planner, but the same procedure
supports correct and safe recovery across all four tested models.

\rqanswer{\textbf{Answer to RQ3.}
StateWise supports correct and safe task execution under both
erroneous target values and defective supporting context.
The unchanged recovery procedure also operates across four
planning models, achieving 80.0\%--94.4\% OC with no safety violations across the episodes.}

\subsection{RQ4: Component Contributions}
\label{sec:rq4}

\begin{wrapfigure}[9]{r}{0.65\linewidth}
  \centering
  \captionsetup{font=scriptsize,skip=3pt}
  \includegraphics[width=0.48\linewidth]{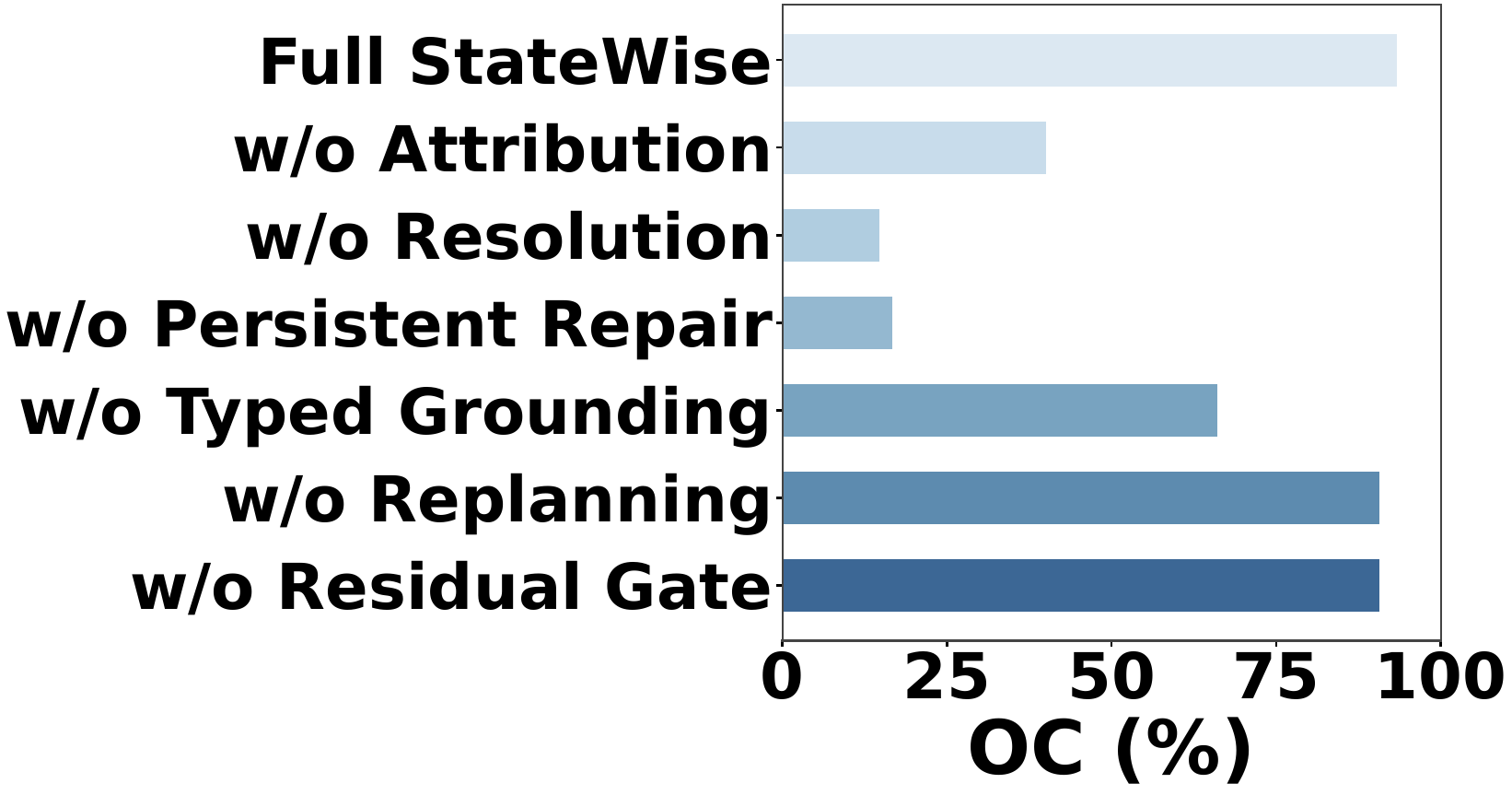}\hfill
  \includegraphics[width=0.48\linewidth]{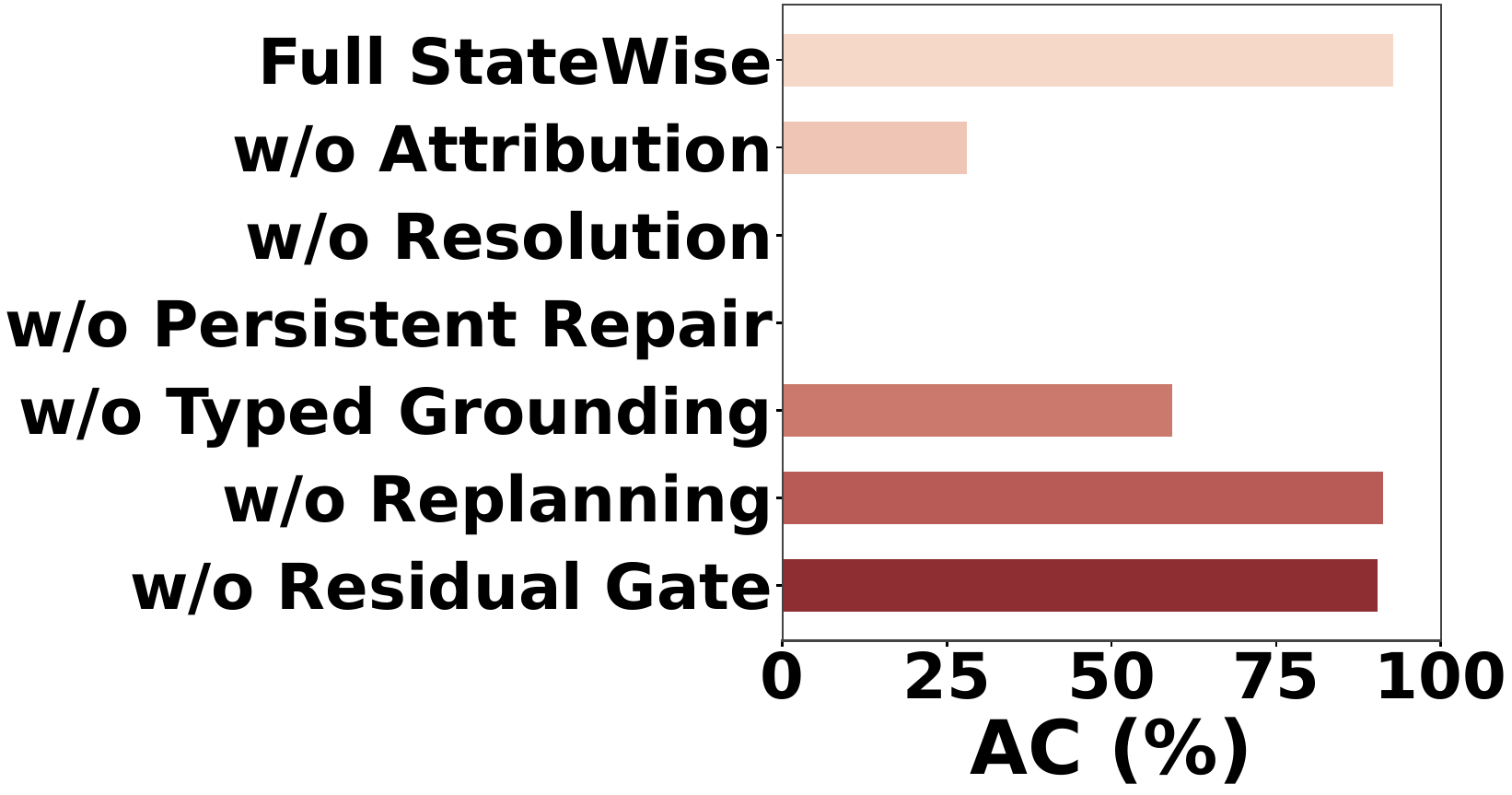}
  \caption{OC and AC of the full StateWise pipeline and six
  ablation variants on the same 150 cases.}
  \Description{Two bar charts compare OC on the left and AC
  on the right for the full StateWise pipeline and six
  component-removal variants.}
  \label{fig:ablation}
\end{wrapfigure}

To understand how StateWise achieves recovery, we remove one
component at a time on the same 150 primary cases.
Figure~\ref{fig:ablation} compares the resulting OC and AC
with those of the full pipeline.

Removing \textit{Resolution} or \textit{Persistent Repair} lowers OC by 78.7
and 76.7 percentage points, respectively, and reduces AC to zero.
These losses highlight the importance of establishing
evidence-supported corrections and making them available to planning,
as described in Sections~\ref{sec:method-resolve}.
Avoiding unsafe execution alone does not restore task completion.

Removing \textit{Attribution} lowers OC by 53.3 percentage points.
Because this variant also disables downstream state intervention,
the loss reflects the combined contribution of that path.
Removing \textit{Typed Grounding} lowers OC by 27.3 percentage points,
showing the importance of binding evidence to the appropriate
record type and action scope.
The losses from these four ablations remain significant
after Holm correction across the six comparisons.

Removing \textit{Replanning} or the \textit{Residual Gate} lowers OC by 2.7
percentage points each.
These stages complete the recovery process by translating repaired
state into a revised action and checking that action before execution.
Their smaller effects complement the larger gains from premise
verification and repair, indicating that the main recovery benefit
comes from correcting the state on which the agent plans.

\rqanswer{\textbf{Answer to RQ4.}
The ablations support StateWise's integrated recovery pipeline:
\textit{Attribution} locates influential premises, \textit{Resolution} and \textit{Typed
Grounding} establish applicable evidence, \textit{Persistent Repair} preserves
corrections, and \textit{Replanning} and the \textit{Residual Gate} turn repaired state
into a checked action.}

\subsection{RQ5: Cross-Task Persistence}
\label{sec:rq5}

We examine whether a correction made by StateWise continues to
support subsequent tasks using 32 four-task sequences.
T1 can trigger repair, while T2--T4 test correction reuse.
Between tasks, conversation and workspace state are reset under
the boundary in Section~\ref{sec:persistence-definition},
with only the persistent store retained.
Table~\ref{tab:rq5} reports task outcomes and AFS.

\begin{wraptable}{l}{0.50\linewidth}
  \centering
  \captionsetup{font=scriptsize,skip=3pt}
  \caption{Correct and safe task outcomes and AFS across
  four-task sequences (\%).}
  \label{tab:rq5}
  \footnotesize
  \setlength{\tabcolsep}{3pt}
  \renewcommand{\arraystretch}{1.1}
  \begin{tabular*}{\linewidth}{@{\extracolsep{\fill}}lrrr@{}}
    \toprule
    Method & T1 & T2--T4 & AFS \\
    \midrule
    Correct Memory & 93.8 & 91.7 & 87.5 \\
    Corrupted Memory & 0.0 & 1.0 & 0.0 \\
    No Memory & 0.0 & 0.0 & 0.0 \\
    MemLineage & 0.0 & 1.0 & 0.0 \\
    Safety Sentry & 0.0 & 2.1 & 0.0 \\
    Ask-or-Assume & 9.4 & 6.2 & 3.1 \\
    \textbf{StateWise} & \textbf{90.6} & \textbf{92.7}
      & \textbf{90.6} \\
    \bottomrule
  \end{tabular*}
\end{wraptable}

Starting from erroneous memory, StateWise achieves 92.7\%
correct and safe completion across subsequent tasks and
90.6\% AFS, reaching the level of the correct-memory reference,
which achieves 91.7\% and 87.5\%, respectively.
RR reaches 97.9\%, showing that repaired records remain
available to later planning even after transient context is cleared.

The tested baselines do not explicitly preserve verified corrections
for subsequent reuse. Their later-task completion rates remain
at 1.0\%--6.2\%, with AFS no higher than 3.1\%, offering
little sustained improvement over corrupted memory.

\rqanswer{\textbf{Answer to RQ5.}
Starting from erroneous memory, StateWise achieves
\textbf{90.6\% AFS} and 92.7\% correct and safe completion
across subsequent tasks, reaching the performance level
of the correct-memory reference.
Persistent repair allows the benefit of an initial correction
to carry forward across the evaluated task sequence.}

\subsection{RQ6: Transfer Across Interfaces}
\label{sec:rq6}

We examine whether StateWise maintains its recovery performance
on repositories, tool operations, and state types not encountered
during development.
After finalizing the recovery method on 40 development cases, we identify
both seen and unseen configurations within the 150-case evaluation set
and compare their respective recovery outcomes using consistent
evaluation criteria.
\begin{wrapfigure}{r}{0.50\linewidth}
  \centering
  \captionsetup{font=scriptsize,skip=3pt}
  \captionsetup[subfigure]{font=scriptsize,skip=2pt}
  \begin{subfigure}[t]{0.48\linewidth}
    \includegraphics[width=\linewidth]{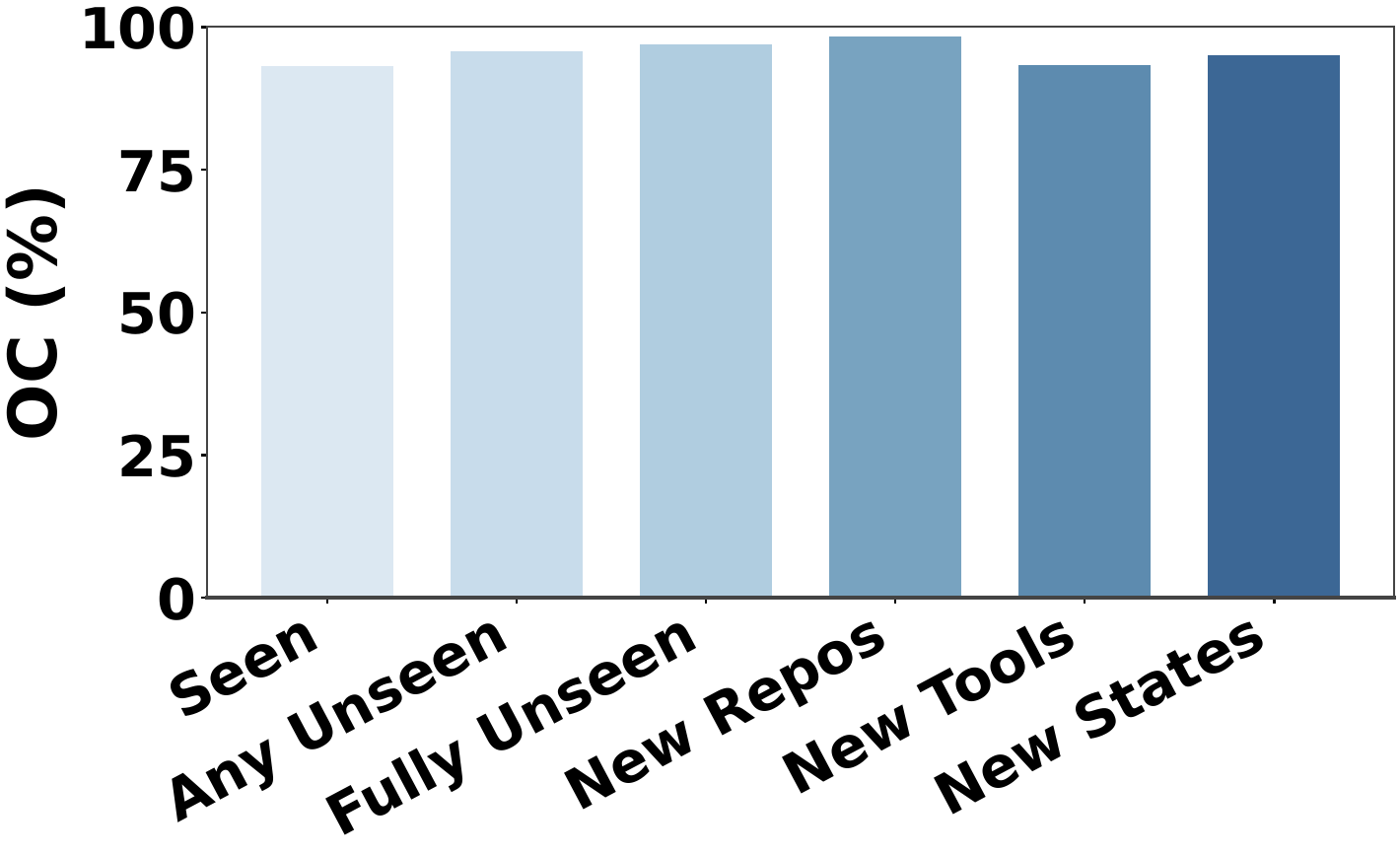}
    \caption{OC}\label{fig:rq6-correctness}
  \end{subfigure}\hfill
  \begin{subfigure}[t]{0.48\linewidth}
    \includegraphics[width=\linewidth]{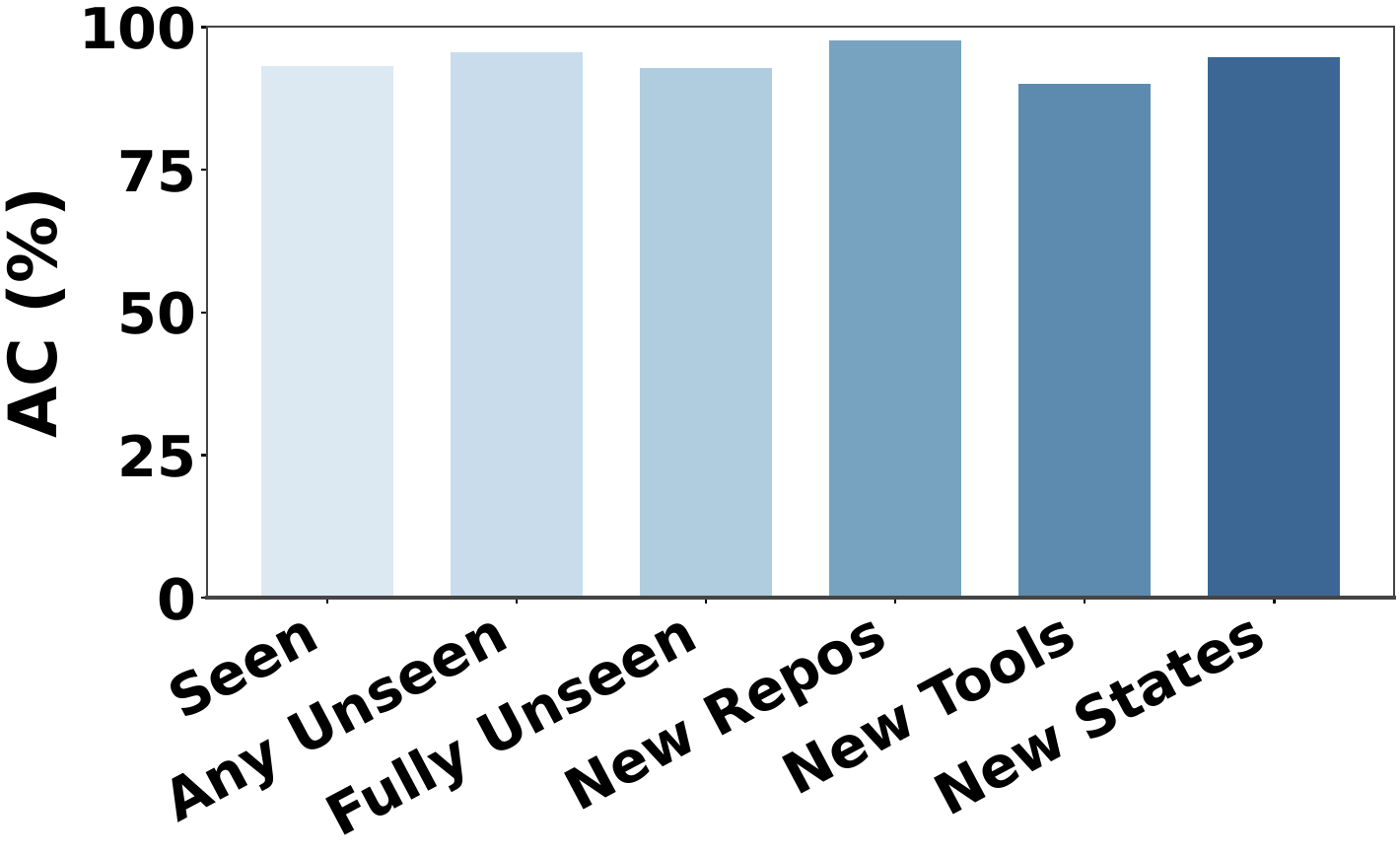}
    \caption{AC}\label{fig:rq6-completion}
  \end{subfigure}
  \par\smallskip
  \begin{subfigure}[t]{0.48\linewidth}
    \includegraphics[width=\linewidth]{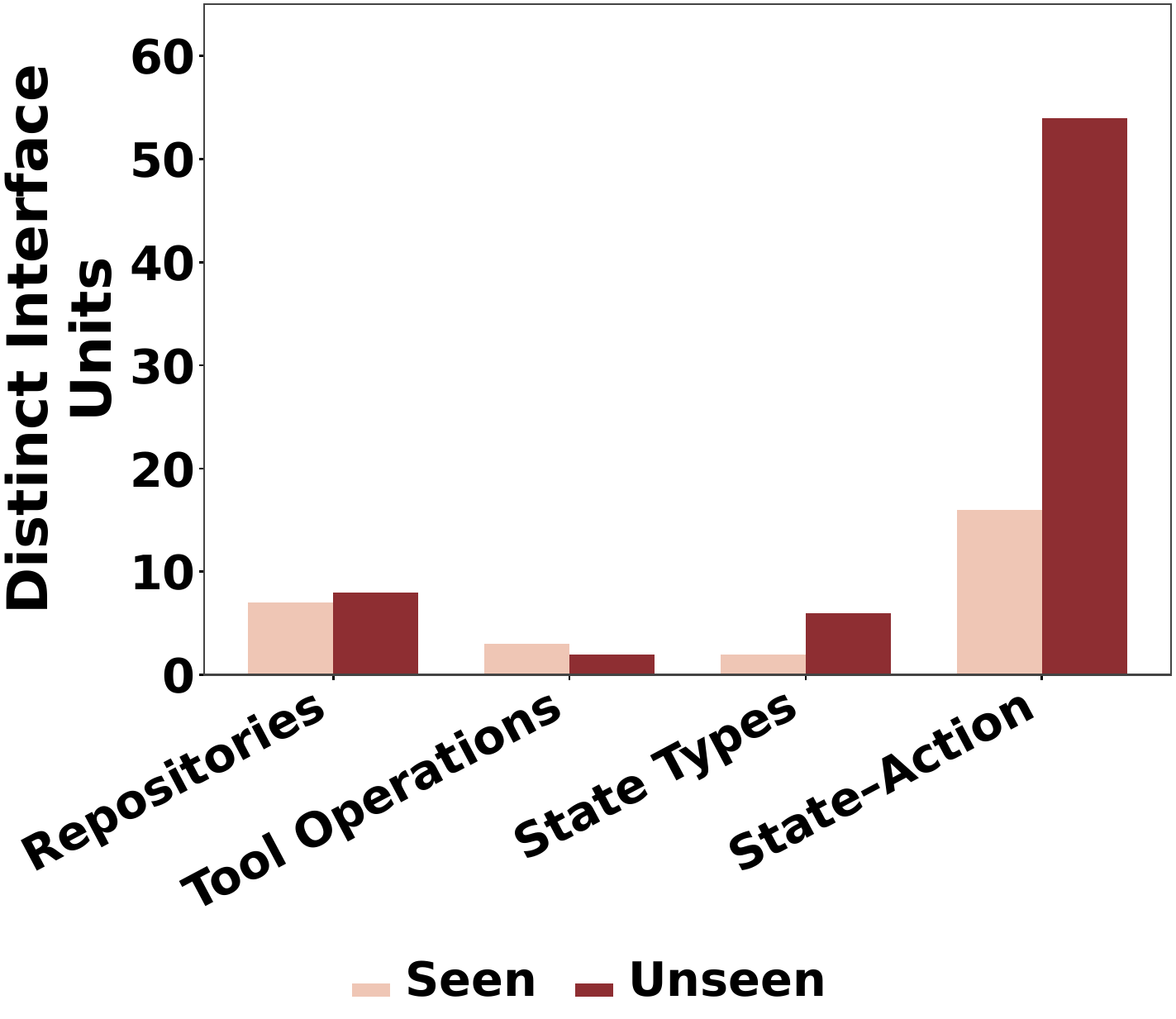}
    \caption{Interface Coverage}\label{fig:rq6-coverage}
  \end{subfigure}\hfill
  \begin{subfigure}[t]{0.48\linewidth}
    \includegraphics[width=\linewidth]{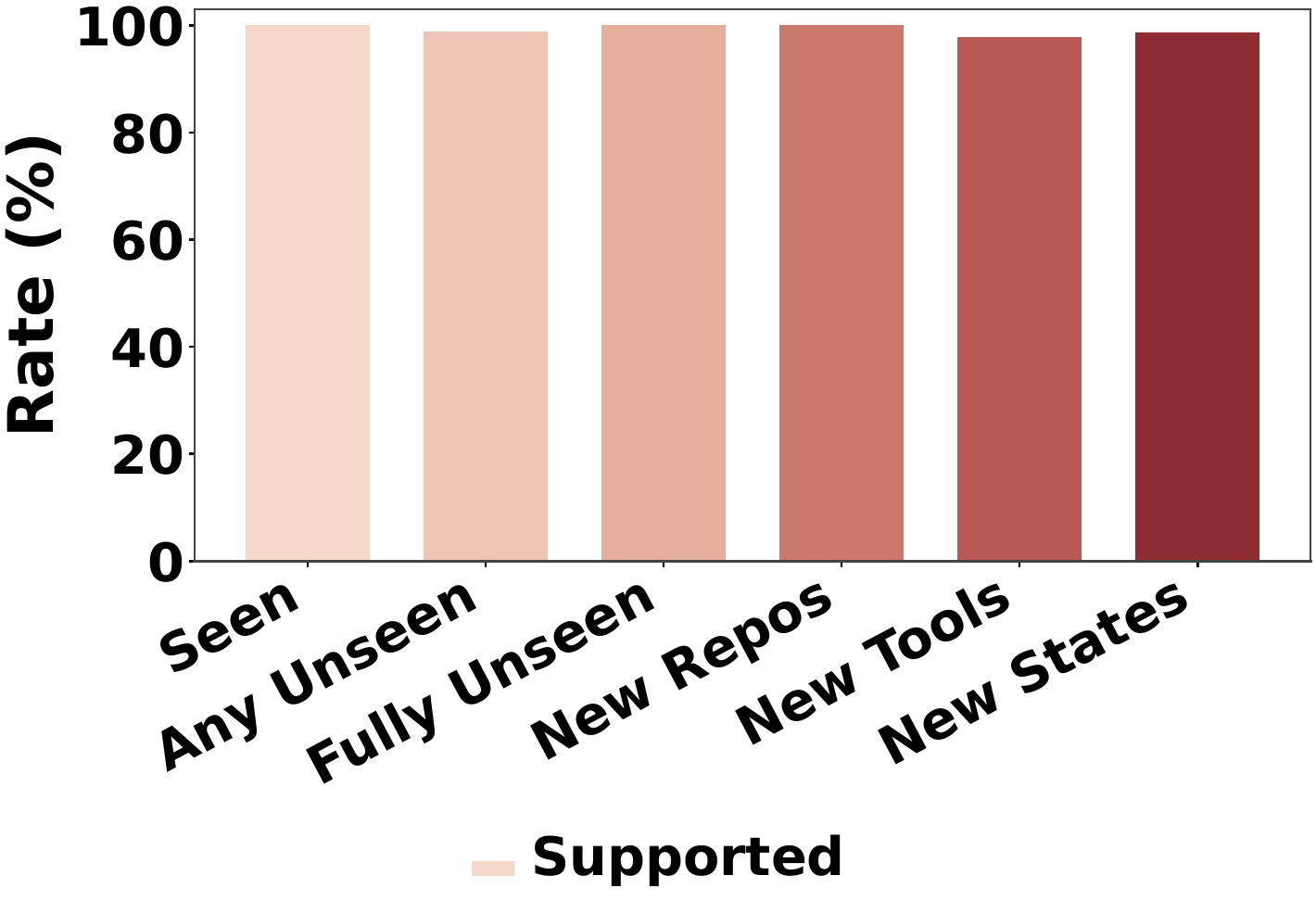}
    \caption{Support and Safety}\label{fig:rq6-support}
  \end{subfigure}
  \caption{Recovery across seen and unseen configurations
  with an unchanged StateWise core and supplied interface contracts.
  Panels report OC, AC, distinct interface units,
  contract support, and UAR. }
  \Description{Four panels summarize StateWise's OC, AC,
  interface coverage, contract support, and UAR across
  seen and unseen configurations.}
  \label{fig:rq6}
\end{wrapfigure}
Figure~\ref{fig:rq6} summarizes the configuration differences
and transfer results.

The evaluation introduces 8 repositories, 2 tool operations,
6 semantic state types, and 54 state--action combinations
not encountered during development.
Each combination comprises a state key, tool operation,
scope, and verifier source.
Of the 150 cases, 106 contain unseen configurations,
including new individual units.

On these 106 cases, StateWise achieves 93.4\% OC and 92.6\% AC,
with zero UAR, compared with 93.2\% OC on seen configurations.
It also achieves 95.1\% OC on cases involving new state types
and 98.4\% on those involving new repositories.
These results indicate that recovery remains effective across
the tested interface changes, allowing the same procedure
to be reused through the corresponding interface contracts.

\rqanswer{\textbf{Answer to RQ6.}
StateWise achieves 93.4\% OC and 92.6\% AC on unseen
configurations, with zero UAR and OC comparable to that
on seen configurations.
These results support reuse across the tested repositories,
tool operations, and state types through supplied interface
contracts, without modifying the core recovery procedure.}

\section{Discussion}
\label{sec:discussion}
The evaluation results motivate treating persistent-state recovery as part
of reliable agent execution.
We discuss how its benefits extend across successive tasks, how \statewise{}
connects record-level action dependence with evidence-supported persistent
updates, and how verification and scope requirements shape its practical
applicability.

\noindent\textbf{State Validity and Task Continuity.}
The paired tests show that a single erroneous record can impair both
task correctness and action safety.
The baseline comparisons distinguish suppressing unsafe operations
from restoring task capability, while ablations identify validity
resolution, typed grounding, and persistent repair as major contributors
to recovery.
The four-task sequences further show that confirmed corrections
remain useful after conversation and workspace resets.
This reset boundary makes persistent memory the channel through which
corrections reach subsequent planning contexts.
These findings motivate treating operational-state maintenance as
part of reliable execution: applicable premises support the current
task, and persistent corrections preserve their usefulness for
subsequent decisions.

\noindent\textbf{Connecting Actions, Evidence, and Persistent Updates.}
\statewise{}'s technical contribution is the correspondence it maintains
among an action's supporting records, evidence of their current validity,
and scoped persistent corrections.
Constrained counterfactual replanning provides evidence of decision
dependence, making the proposed action a focused target for state
diagnosis.
The record's type and authority then determine how its validity is
resolved, and typed evidence grounding binds the resulting correction
to the same record and its established scope.
This binding specifies which active value is superseded and which
unrelated records must remain unchanged.
Persistent writeback retains this correction and its lineage for
current replanning and later retrieval.
This connection turns action-triggered verification into reusable
state maintenance.
An independent state--action check assesses the revised operation
against the verified state and execution constraints.

\noindent\textbf{Evidence and Scope Boundaries.}
Applicability depends on explicit record--action bindings, accessible
authoritative evidence, and established repair scopes for persistent updates.
These requirements fit recurring coding workflows involving inspectable
runtime and workspace state, verifiable repository policies, and
developer-owned choices that can be resolved through targeted clarification.
For example, environment changes and repository policy updates can
invalidate stored records that guide repeated coding tasks.
The transfer results support extending this structure through supplied
interface contracts and corresponding verifier adapters.

The audit relies on model annotation and mainly covers one agent family.
Executable evaluations use controlled single-record defects, simulated
developer responses, a limited cross-model set, and four-task sequences.
Longer deployed studies with independent human annotation should examine
interacting records, conflicting corrections, and partial evidence.
Extension to external services and collaborative agents additionally
requires broader verifier coverage and mechanisms for resolving changing
authority and overlapping scopes.

\section{Conclusion}
\label{sec:conclusion}

In this work, we studied how invalid persistent operational records influence
coding-agent decisions and proposed \statewise{} for pre-action
diagnosis and recovery.
The framework connects action attribution, current-validity verification,
and scoped persistent repair with replanning and independent action
checks, preserving confirmed corrections for later tasks.
On the primary 150-case benchmark, \statewise{} achieves 93.3\%
overall correctness under corrupted persistent state, compared with
38.7\% for the baseline, with no unsafe actions observed.
This work addresses a gap between intervening in the current action
and correcting the persistent premise that guides subsequent decisions.
An invalid record can survive a blocked operation or a locally corrected
decision and continue to influence later plans.
\statewise{} makes the corresponding record and its applicable scope
explicit targets of evidence-supported recovery.
These findings connect agent-memory research with planning and action
safety through the validity of operational premises.
They motivate evaluating operational memory through its current
applicability, influence on actions, and reuse of verified corrections,
and integrating state maintenance into the execution architecture
of agents operating across tasks and changing environments.

\section*{Data Availability}
\StateWiseDataAvailability

\StateWiseAcknowledgments

\bibliographystyle{ACM-Reference-Format}
\bibliography{references}

\end{document}